\documentclass[usenatbib]{mn2e}

\usepackage[varg]{txfonts}

\usepackage{booktabs}
\usepackage{makeidx}
\usepackage{graphicx}
\usepackage[dvipsnames, usenames]{color}
\usepackage{colortbl}
\usepackage{hyperref}
\usepackage{url}
\usepackage{breakurl}
\usepackage{natbib}
\bibpunct{(}{)}{;}{a}{}{,}

\title[New hot subdwarf stars by means of VO tools II]{A search for new hot subdwarf stars by means of Virtual Observatory tools II}
\author[E. P\' erez-Fern\'andez, A. Ulla, E. Solano et al.]{E. P\'erez-Fern\'andez $^{1,2}$\thanks{E-mail:
estherperez@edu.xunta.es}, A. Ulla$^{2}$, E. Solano$^{3,4}$, R. Oreiro$^{5}$ and C. Rodrigo$^{3,4}$\\
$^{1}$ IES de Beade, Conseller\'\i a de Educaci\'on e O.U., Cami\~no de Outeiro 10, 36312 Vigo, Spain \\
$^{2}$Departamento de F\'\i sica Aplicada, Universidade de Vigo, Campus Lagoas-Marcosende, 36310 Vigo, Spain\\
$^{3}$Departamento de Astrof\'\i sica, Centro de Astrobiolog\'\i a (INTA-CSIC), PO Box 78, E-28691 Villanueva de la Ca\~nada (Madrid) \\
$^{4}$ Spanish Virtual Observatory \\
$^{5}$ Instituto de Astrof\'\i sica de Andaluc\'\i a (IAA-CSIC), Glorieta de la Astronom\'\i a, s/n, 18008 Granada.}
\begin{document}

\date{Accepted 2016 January 21;  Received 2016 January 20; in original form 2015 July 6}

\pagerange{\pageref{firstpage}--\pageref{lastpage}} \pubyear{2016}

\maketitle

\label{firstpage}

\begin{abstract}
Recent massive sky surveys in different bandwidths are providing new opportunities to modern astronomy. The Virtual Observatory (VO) represents the adequate framework to handle the huge amount of information available and filter out data according to specific requirements.

In this work, we applied a selection strategy to find new, uncatalogued hot subdwarfs making use of VO tools. We used large area catalogues (GALEX, SDSS, SuperCosmos, 2MASS) to retrieve photometric and astrometric information of stellar objects. To these objects, we applied colour and proper motion filters, together with an effective temperature cutoff, aimed at separating hot subdwarfs from other blue objects such as white dwarfs, cataclysmic variables or main sequence OB stars.  As a result, we obtained 437 new, uncatalogued hot subdwarf candidates. Based on previous results, we expect our procedure to have an overall efficiency of at least 80 per cent. Visual inspection of the 68 candidates with SDSS spectrum showed that 65 can be classified as hot subdwarfs: 5 sdOs, 25 sdOBs and 35 sdBs. This success rate above 95 per cent proves the robustness and efficiency of our methodology. 

The spectral energy distribution of 45 per cent of the subdwarf candidates showed infrared excesses, a signature of their probable binary nature. The stellar companions of the binary systems so detected are expected to be late-type main sequence stars. A detailed determination of temperatures and spectral classification of the cool companions will be presented in a forthcoming work.

\end{abstract}


\section{Introduction}

Hot subdwarf (hot sd) stars are core-helium burning stars at the end of the horizontal branch or even beyond that stage. The origin of these faint, blue objects is still a matter of controversy. With effective temperatures exceeding 19000 K and logg $\geq$ 5, hot sds are objects that have lost  most of their H envelope in previous evolutionary stages, leading to a $\sim 0.5M_{\sun}$ star. They are unable to follow canonical evolution through the Asymptotic Giant Branch (AGB) proceeding, instead, directly towards the white dwarf cooling track. Circumstances that lead to the removal of all but a tiny fraction of the hydrogen envelope, at about the same time as the core has achieved the mass required for the He flash ($\sim 0.5 M_{\sun}$ ), are still a matter of debate. Theoretical evolution scenarios proposed so far include enhancement of the mass loss efficiency near the red giant branch (RGB) tip \citep{Cruz96} or mass transfer through binary interaction \citep{Mengel96}. See \cite{Heber_rev} for a review on observational and theoretical aspects of hot subdwarfs, or \cite{Geier13b} for more recent discoveries.

Hot subdwarfs are found in the field, both in the disk and halo, but also populating the most Extreme part of the Horizontal Branch (EHB) of some Galactic clusters. Based on this observational evidence,  they have been proposed as an explanation for the UV-upturn phenomenon shown in some elliptical galaxies \citep{Brown1997}.

Hot sds are divided in two main classes, sdBs and sdOs, according to composition. SdB spectra are dominated by the Balmer series, while sdOs are hotter objects caracterized by the presence of He\,{\sc ii} 4686\AA \,  and the Pickering series.  Additionally, a variety of He\,{\sc i} lines may appear in both classes, and some sdOs show metallic C or N lines. More complex classification schemes have been proposed in  \cite*{PGcat} or more recently in \cite{Drilling13}.

The subdwarf database \citep{Roydatabase} catalogues 1600  sdBs and 500 sdOs spectroscopically confirmed hot subdwarfs.  A significant number of new hot sds have been discovered in more recent studies like \cite{Vennes2011},   \cite{MUCHFUSS, Geier15_RVcat}, \cite{Nemeth2012} , \cite{Kleinman}, \cite{Kupfer2015}  or \cite{Kepler15}.  Increasing the number of hot sds is important for a robust statistical confrontation with theoretical evolutionary scenarios. It may also lead to the discovery of interesting objects that are still scarce, such as pulsating sdBs  or sdOs \citep{pulsating}, eclipsing or reflecting hot sd binaries \citep{For2010, Derekas15}, and hot sds as central stars of planetary nebulae \citep{Alba15}. All of them are particularly interesting  for studying the stellar interiors, the mass transfer mechanism at work, as well as the evolutionary formation channels, and would contribute to better understand these evolved objects.

Hot subdwarfs were first found analyzing faint blue stars, starting with the  \cite{Humason47}  survey or the Palomar-Green (PG) catalogue \citep{PGcat}. At present we have at our disposal deeper and more extensive surveys, covering large regions of the sky and wide spectral ranges. Besides, with online access tools like the Virtual Observatory (VO)\footnote{\url{http://www.ivoa.net/}} we can access  data from most of these surveys in a very efficient way, crossmatching information to select  objects with particular characteristics.

In this regard, the aim of the work here presented is to obtain a number of new hot sd candidates as large as possible. We apply the selection process developed in \cite{Oreiro2011} (hereafter paper I), that combines photometric and proper motion information from different surveys, making use of VO tools, with the intention of discriminating hot sds from other types of objects  of similar colours, mainly white dwarfs (WD), cataclysmic variables (CV) and main-sequence O and B stars, considered as contamination sources in this work.

Section 2  describes the methodology employed, Section 3 the results obtained together with their analysis, Section 4 a summary of our main achievements on spectral classification of the new hot sd candidates discovered, and Section 5 ends with a general summary and conclusions.

\section{Methodology}

The methodology described in paper I is reproduced  here. In that work, a hot sd selection procedure was defined and tested by means of a thorough retrieval, with the aid of VO tools, of multi-colour photometry and astrometric information from stellar catalogues. A filtering procedure to distinguish among different types of objects was designed  to obtain a hot subdwarf sample with a low contamination factor. The method was tested on two sky regions: the Kepler FoV\footnote{\url{http://kepler.nasa.gov/science/about/targetFieldOfView/} } and a region of 300 $deg^2$ around ($\alpha$:225, $\delta$:5 deg)  obtaining a high rate of success (above $80$ per cent) in finding new uncatalogued hot subdwarfs. Temperatures were provided by fitting their spectral energy distribution (SED), and considering two-atmosphere fits for those objects with a clear infrared excess, a signature of the possible presence of a cool companion. 

\begin{figure}
\resizebox{\hsize}{!}{\includegraphics{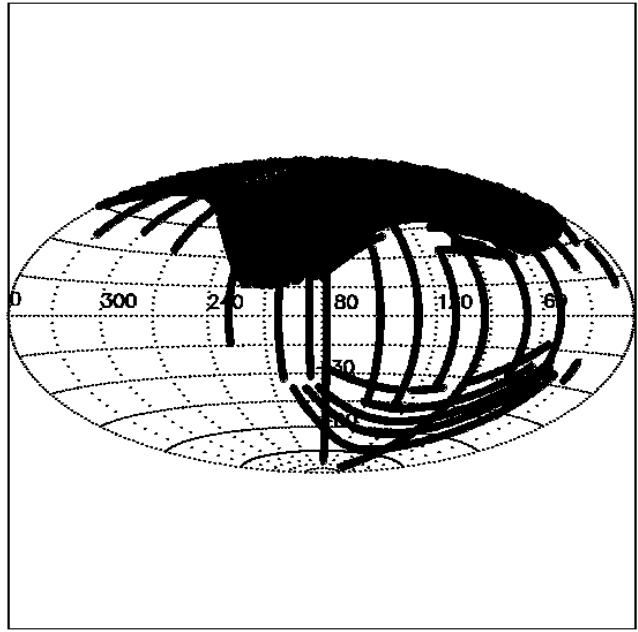}}
\caption{ SDSS DR7 imaging coverage in Galactic coordinates, which is the region covered by this study. Source: \url{http://classic.sdss.org/dr7/coverage/}}
\label{dr7}
\end{figure}

 Once tested the validity of the proposed strategy, in this work we apply it to a wider sky region. The extension of this region is about 11663$deg^2$, limited by the SDSS\footnote{\url{http://www.sdss.org/}} DR7 survey coverage \citep{DR7}. Most of the Galactic northern cap down to b=+30 is covered, as well as some strip-shaped areas crossing the Galactic plane and reaching the southern Galactic hemisphere. Fig. \ref{dr7} shows the footprint in Galactic coordinates of the regions studied.

We remind the reader that the principal aim of this work is to apply a reliable selection procedure to find a large number of new hot subdwarf candidates, and not to perform a deep analysis of each star physical parameters. In this regard, we computed effective temperature estimates using an automated procedure provided by VOSA \citep{VOSA}, a very useful online facility of the Spanish Virtual Observatory\footnote{\url{http://svo.cab.inta-csic.es}}.  
 
For the ease of the reader, we outline here the selection procedure.

\begin{itemize}
\item {\bf Hot subdwarf selection filters:} First step was to crossmatch the photometric and proper motion surveys and apply to the retained sources the cuts aimed at selecting hot subdwarfs. 

The surveys employed were SDSS DR7, GALEX GR6/GR7\footnote{\url{http://galex.stsci.edu/}} \citep{Galex}, 2MASS Point Source Catalogue\footnote{\url{http://www.ipac.caltech.edu/2mass/}} \citep{2MASS} and SuperCosmos\footnote{\url{http://surveys.roe.ac.uk/ssa/}} \citep{SuperCosmos}. We required the GALEX sources to have measured magnitudes in both filters ($FUV>0$, $NUV>0$) and to be brighter than  5$\sigma$ of the magnitude limit ($FUV<19.9$, $NUV<20.8$). Sources must also be classified as point objects by SDSS (cl=6).  We retained sources with counterparts in all the surveys within a maximum distance of 5 arcsec.
To the selected sources we applied the following cuts, as discussed in paper I: 
\begin{eqnarray}
&& -4<(FUV_0-Ks_0) <0.5 \label{filter1} \\
&&-2 <(FUV_0 - NUV_0)<0.2 \label{filter2} \\
&& 19 < H(NUV_0) < 27 \label{filter3}
\end{eqnarray}
where the 0 subscript stands for Galactic extinction corrected magnitudes, and $H(NUV_0)$ for the reduced proper motion of the $NUV$ filter. 

For bright stars, severe calibration problems in the GALEX photometry have been pointed out by \cite{Camarota2014} who, using a well studied sample of WDs with UV spectra, derived empirical corrections to the GALEX magnitudes in the non-linear range. The corrections are valid within the $9.321 < NUV < 17.5$ or $10.509 < FUV<17.5$ ranges. We have thus identified the stars in our sample lying within those limits, and applied to them the correction factors established in that paper.

\item {\bf Discriminating new from already classified objects:} We crossmatched our list with published and well-established catalogues of spectroscopically confirmed subdwarfs, white dwarfs, cataclysmic variables and OB stars. These include: 
\begin{itemize}
\item {\sl The subdwarf database for hot sds} \citep{Roydatabase}. 
\item {\sl A selection of hot subluminous stars in the Galex survey} \citep{Vennes2011, Nemeth2012}
\item {\sl The photometric and spectroscopic catalogue for luminous stars}  \citep{Reed}.
\item {\sl The catalogue of Cataclysmic Variables}, version 2006 \citep{Downes}.
\item {\sl The SDSS DR7 white dwarf catalogue} \citep{Kleinman}.
\item {\sl A Catalogue of Spectroscopically Identified White Dwarfs}, version 2008 \citep{Cook}
\end{itemize}

Sources already available in these catalogues were discarded. The remaining objects were searched in SIMBAD\footnote{\url{http://simbad.u-strasbg.fr/simbad/}}, VIZIER.\footnote{\url{http://vizier.u-strasbg.fr/viz-bin/VizieR}} and any catalogue available through online VO tools.   Very recent catalogues like \cite{Geier15_RVcat}, \cite{Kupfer2015} and \cite{Gentile2015} were considered in the Vizier search. Other catalogues containing spectroscopically confirmed hot subdwarfs but not included in Vizier \citep{Vennes2011, Nemeth2012, Kawka2015, Kepler15} were also inspected. Any source already spectroscopically classified in these catalogues was discarded.

\item {\bf Spectral distribution fit:} For each object in the pre-candidate list, we used VOSA to accomplish the following steps:
\begin{itemize}
\item Gathering of additional photometry: GALEX-SDSS-2MASS photometry was complemented with additional photometry from UKIDSS\footnote{\url{http://surveys.roe.ac.uk/wsa/}} LAS DR9 \citep{UKIDSS}, Tycho-2 \citep{Tycho} and WISE\footnote{\url{http://wise.ssl.berkeley.edu}} \citep{WISE}. Some candidates had saturated or bad SDSS photometry. In these cases we replaced SDSS by UCAC4 \citep{UCAC4} photometry, if available.
\item Magnitude-to-flux transformation: VOSA used the gathered photometric information to calculate the absolute fluxes and their associated errors taking advantage of the Filter Profile Service\footnote{\url{http://svo2.cab.inta-csic.es/theory/fps3/}} (FPS), a service developed by the Spanish Virtual Observatory to provide VO access and representation of many of the most common photometric systems in astrophysics. Fluxes were then dereddened using the extinction law by \cite{Fitz99} and the $E(B-V)$ values available in the GALEX catalogue, which in turn have been taken from the \cite*{Schlegel} extinction maps. 
\item Model comparison: The flux-dereddened observational SEDs were then compared to the TLUSTY OSTAR2002+BSTAR2006 NTLE models for O and B stars \citep{Hubeny1995,Lanz2003,Lanz2007} implemented at VOSA to derive effective temperatures. We considered the whole model grid, with $T_{\rm eff}$ ranging from 15000 to 55000K. In the SEDs’ fitting procedure both surface gravity and metallicity were simply left as free parameters, as their impact on the effective temperature determination can be considered as negligible. Therefore, we warn the reader that the gravity and metallicity values obtained from the SED fitting cannot be considered as the real physical parameters of the objects listed in Tables \ref{table:1} - \ref{table:4} below. 

\cite{Heber2000}  have shown that the use of LTE vs NLTE model atmospheres yields almost identical $T_{\rm eff}$'s and only systematic $\log g$ differences, at least when fitting hot sd spectral lines. We do not expect other result in our procedure of fitting SEDs.

We have performed, anyway, a comparison between the effective temperatures calculated using the TLUSTY and Kurucz \citep{Castelli97} grids of atmospheric models. Only objects with a good SED fit flag (5XX) and a TLUSTY temperature value lower than 35000K (to avoid boundary problems with the maximum temperature of the grids) were considered. We obtained a difference in effective temperatures below $10$ per cent for $90$ per cent of the objects (or $82$ per cent of objects for an up to $5$ per cent difference), indicating  that, as expected,  the NLTE effects on $T_{\rm eff}$ determination can be neglected.

We also attempted to leave $A_v$ as a free parameter in the SED fitting process. Nevertheless, due to the $A_v-T_{\rm eff}$ degeneracy, this exercise rendered multiple solutions and we finally decided to include extinction as a fixed parameter. 
\end{itemize}

\item {\bf Source image checking:}  Finally, we visually inspected using Aladin\footnote{\url{http://aladin.u-strasbg.fr/}} the SDSS images and catalogue data of our pre-candidate list of targets to discard instrumental features, bad crossmatches or contamination from nearby, bright sources.
 
 In fact, we found some cases with a clear mismatch between GALEX, SDSS and 2MASS sources. These pathological cases are mostly due to the different spectral coverage and limiting magnitude of the surveys. We kept these objects without infrared photometry in a separate list, as they appear to be very hot and blue objects, and thus interesting from our point of view (Table \ref{table:4}).

\end{itemize}

\section{Results}
\label{sec:results}

After crossmatching the photometric surveys and applying the selection filters in equations (\ref{filter1})-(\ref{filter3}), we ended up with a list of 1242 pre-candidates. 638 of them were already classified in the literature, with the following percentages: $83$ per cent hot subdwarfs, $12$ per cent WD, $2$ per cent CV, $2$ per cent B stars, and less than $0.5$ per cent other main sequence stars. These numbers agree with those obtained in paper I, demonstrating the robustness of our selection procedure. The remaining 604 pre-candidates where not found in the literature.

\begin{figure}
\resizebox{\hsize}{!}{\includegraphics{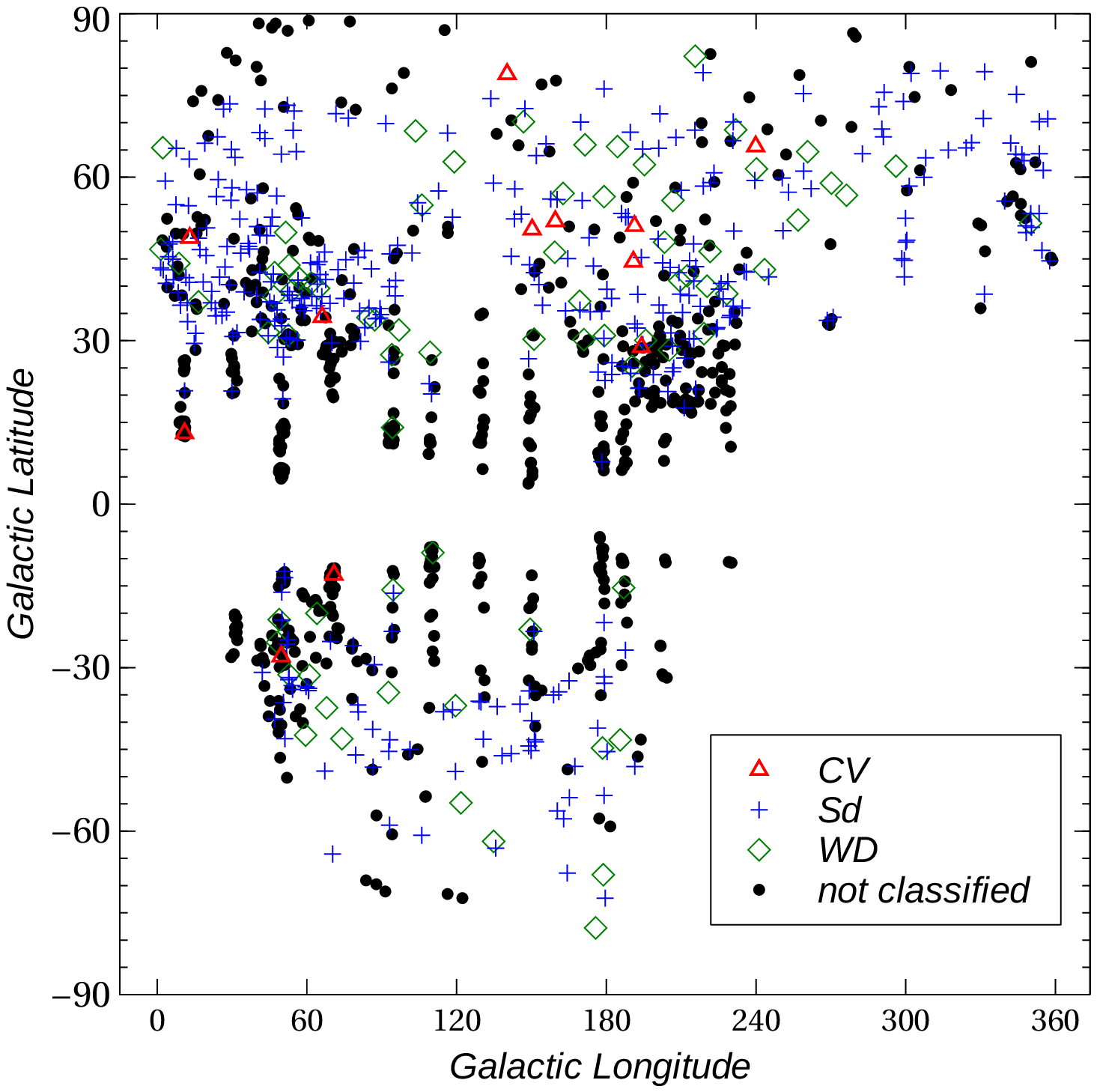}}
\caption{Galactic coordinates of the pre-candidates after the photometric and proper motion cuts. Those not yet classified in the literature are marked with black circles.}
\label{galactic}
\end{figure}

In Fig. \ref{galactic} the classified and unclassified objects selected by the photometric and proper motion cuts are pictured in Galactic coordinates. Notice that a large fraction of the unclassified objects lay in the bands near the Galactic plane, as these tend to be less studied regions.

\subsection{Effective temperatures}
\label{subsec:fit}

Effective temperatures were obtained from the comparison between the observational SEDs and the TLUSTY models. After the fitting, we kept candidates with $T_{\rm eff} > 19000K$, provided that the fit was good. Sources with $T_{\rm eff} < 19000K$ and a bad SED fitting were also kept, as this could be a signal of a binary candidate. 167 out of the 604 unclassified sources did not pass the cut, leaving us with a list of 437 final subdwarf candidates.

The bad fits are of mainly three different sorts: excess in the red part of the spectrum (IR), ultraviolet (UV) excess, and both IR and UV excesses in the same source. IR excesses are probably a signature of a binary system. UV excess could also indicate the presence of a very hot companion, but uncertainties associated to the ultraviolet extinction correction cannot be discarded. $E(B-V)$ values have been taken from \cite{Schlegel}, who seem to overestimate the reddening to lines of sight where $A_V \geq 0.5$ mag \citep{Arce99}.

As a further check, we also performed in VOSA a Bayesian analysis of the model fits. We found that, for 356 sources, the probability associated to the $T_{\rm eff}$ value obtained from the chi-square fitting was over 80 per cent. For the rest of sources (81), we provide an effective temperature interval covering an accumulated probability of, at least, 80 per cent. With this procedure we obtain fairly temperature estimations for the whole sample. The only exception to this were the targets whose temperature estimate reached the upper limit of the TLUSTY models, 55000K. This is not suprising as we know that some sdOs can achieve very high temperatures \citep{Stro2007}. For these objects, just a lower limit in effective temperatures is provided.

To classify the quality of the fits we tagged each target  with a three digits quality flag (see Tables \ref{table:1}-\ref{table:4}). The first digit ranges from one to five: `5' represents good SED fitting, `4' stands for excess in the red part of the SED (IR excess), `3' for both IR and UV excesses, `2' for UV excess only, and `1' for a bad fitting of any other sort. Excesses in the infrared or ultraviolet part of the SED were defined whenever the relative difference between the model and observed (dereddened) values was above $20$ per cent, and the difference increased with decreasing/increasing wavelength, for UV and IR, respectively. We found this criterion matched quite well with a visual inspection of the SED fits. Differences without a clear pattern, in the middle or any part of the SED, were considered bad fits of type `1'.

Second and third digits refer to the quality of the GALEX data: a `1' in the second position represents a problematic GALEX artifact, and a `1' in the third position stands for a bad flag in the photometry extraction\footnote{see the GALEX documentation at \url{http://galex.stsci.edu/GR6/?page=ddfaq\#6}}. In both cases, `0' stands for a good GALEX flag. 2MASS quality flags were  also considered: Photometric values with an $U$ flag ($U$ standing for {\sl upper limit} in magnitude) were not taken into account to perform the SED fitting.

\begin{figure*}
\centering
\resizebox{\hsize}{!}{\includegraphics[width=16cm]{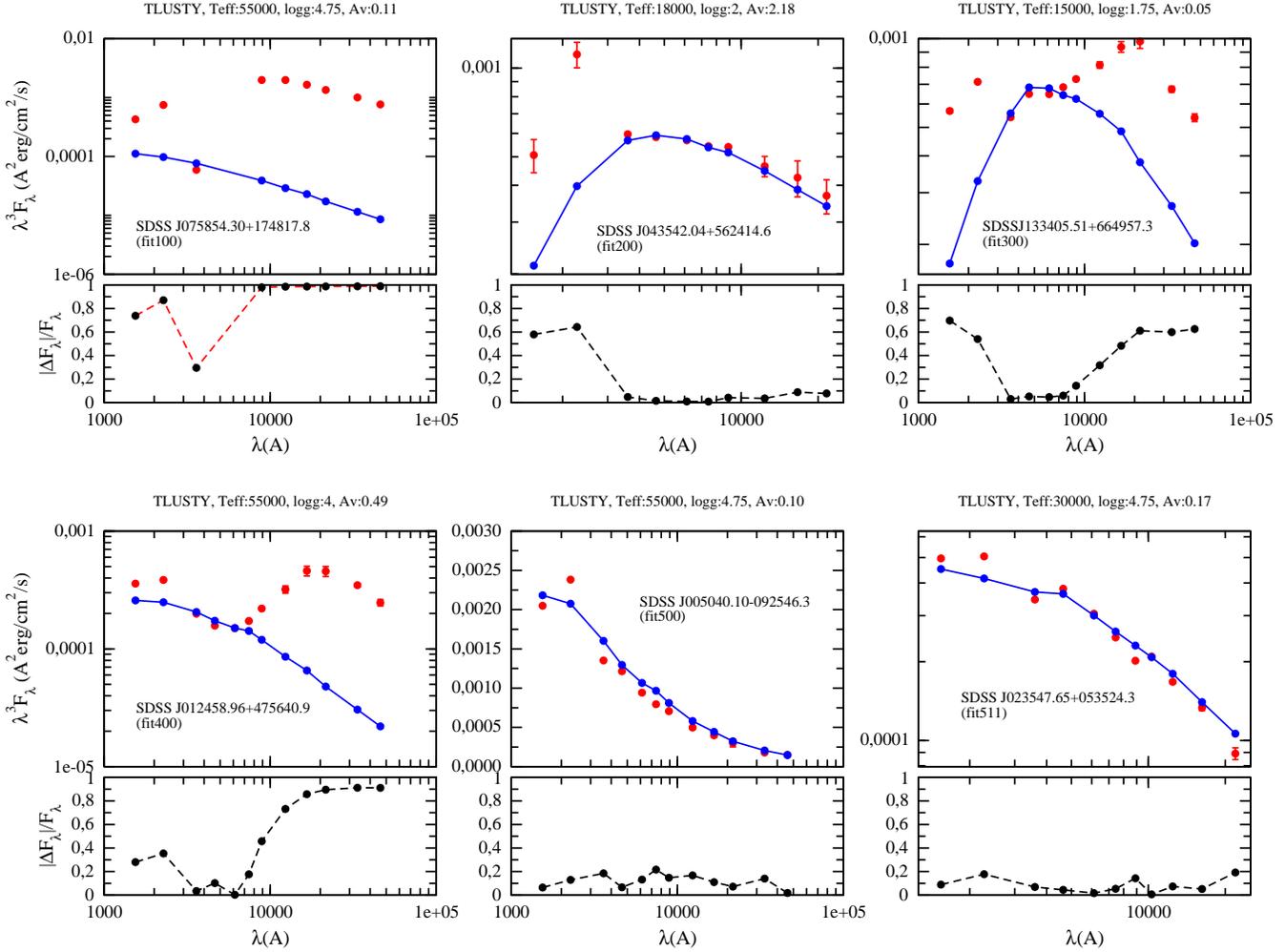}}
\caption{Examples of different quality SED fittings performed by VOSA. From left to right and up to botton: bad fitting (flag 100),  ultraviolet excess (flag 200), UV and IR  excesses (flag 300),  infrared excess (flag 400), good fitting (flag500) and good fitting with bad flags in GALEX phtometry (flag 511). Red points (grey in the grey-scale version of the figure) represent the derredened magnitudes of the given object, connected blue points (dark grey) the magnitudes given by the best fit model. Under each graph we represent the residuals of the observed data and models.}
\label{vosas}
\end{figure*}

Examples of the different quality fits can be seen in Fig. \ref{vosas}. Red points (grey in the grey-scale version of the figure) represent the derredened magnitudes of the given object and the conected blue points (dark grey) the synthetic magnitudes that best fit. Under each graph we represent the residuals of the observed data and models.

In Tables \ref{table:1}-\ref{table:4} we present a sample of the hot subdwarf candidates found by our selection method. Table \ref{table:1} includes good fitted objects with photometric data ranging from the ultraviolet to the infrared, and thus represents clear single candidates (192 objects). Table \ref{table:2} shows sources with excess in the red part of the SED, the most clear binary candidates (110 objects). In Table \ref{table:3} we included the rest of the bad fitted objects (115). Finally, Table \ref{table:4} includes hot objects with no infrared photometry available (20). 

 $T_{\rm eff}$ estimates for candidates with flags 4XX, 3XX, 2XX and 1XX  must be treated with caution, which is warned by means of one of these  bad fit flags.  We remind the reader that, although the estimated $T_{\rm eff}$ are below 19000K, these objects  are kept in the candidate list because the combination of bad fit and low temperature is used as indicator for the presence of binary systems.

In all tables, $FUV$ and $NUV$ were taken from the GALEX archive, and corrected as explained above, if necessary; $u$, $g$, $r$ are from SDSS Data Release 7 and $J$, $H$, $K$ from the 2MASS Point Source Catalogue. We included a column with the 2MASS quality flags of the source. $T_{\rm eff}$ is obtained from the best SED fit performed by VOSA. As explained above, an interval in the temperature column is given whenever the Bayes analysis gave the most likely $T_{\rm eff}$ value with a  probability below $80$ per cent. The {\sl fit flag} column shows our notation for the different qualities in the VOSA $T_{\rm eff}$ fit. In Tables \ref{table:2} and \ref{table:3} we also included the filter where the  excess begins and the expected spectral type of the stellar companion, according to the criterium explained in the next section.

Full tables, with all the photometric filters and other data, including links to the SED fitting diagrams and the SDSS spectrum, when available, can be accessed using the SVO hot subdwarf archive (see Appendix \ref{sec:apen}).

\begin{figure}
\resizebox{\hsize}{!}{\includegraphics{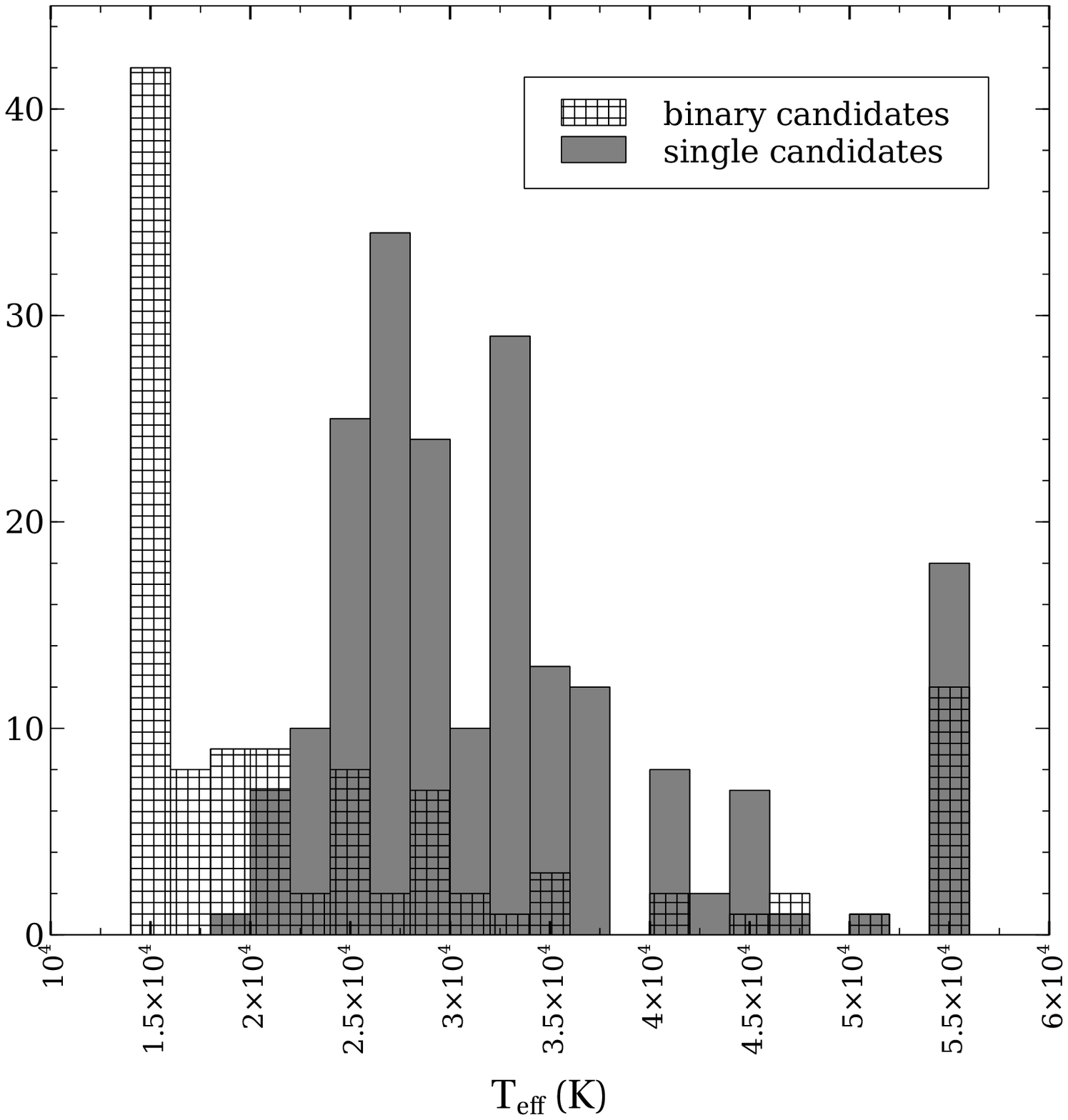}}
\caption{Effective temperature histogram for single (fit flag 5XX) and binary (fit flag 4XX) candidates. The two peaks at 15000 and 55000K show the boundary effects of the TLUSTY models limits (see \ref{subsec:fit} for more details).}
\label{histogram}
\end{figure}

Fig. \ref{histogram} shows a histogram of the effective temperatures obtained for both the good fitted single candidates (fit flag 5XX) and the most clear binary candidates (fit flag 4XX). The majority of stars within the single sample lay in the temperature range 20000-30000K while the effective temperatures of the binary sample are shifted towards lower values. This is not surprising, as we are including in the 4XX category objects with effective temperatures below 19000 K (see above). On the other hand, the peaks at 15000K and 55000K are signaling the limits in $T_{\rm eff}$ of the TLUSTY grid of models.

\begin{table*}
\caption{A sample of subdwarf candidates with good SED fit. $FUV$ and $NUV$ were taken from the GALEX archive (and corrected as explained in the text, if necessary); $u$, $g$, $r$ are from SDSS Data Release 7 and $J$, $H$, $K$ from the 2MASS Point Source Catalogue. We included the 2MASS quality flags of the source, where `U' stands for upper limit in the corresponding photometric value. $T_{\rm eff}$ is obtained from the best SED fit performed by VOSA. An interval in the temperature column is given whenever the Bayes analysis gave the most likely $T_{\rm eff}$ value with a  probability below $80$ per cent. The last column represents a quality flag on the $T_{\rm eff}$ fit: `5' in the first digit stands for good fitting; a `1' in the second or third digit represents some problem in the GALEX photometry. The complete table can be found at \url{http://svo2.cab.inta-csic.es/vocats/hsa/}.}
% title of Table
\label{table:1}
% is used to refer this table in the text
\centering
% used for centering table
\begin{tabular}{c c c  c c c c c c c c c c  }
% centered columns (13 columns)
\hline\hline
% inserts double horizontal lines
RA  & DEC   & NUV   & FUV       & u        & g       & r       & J       & H        & K        & 2MASS   & $T_{\rm eff}$    & Fit     \\
(J2000)   & (J2000)   &      &   & & & & & & & flag    &   (VOSA) &    flag     \\
% table heading
\hline
% inserts single horizontal line
 00:03:07   & +24:12:12   & 16.028    & 15.916  & 16.105   & 16.148   & 16.533   & 16.628   & 16.085   & 17.106   & BUU          & 25000   & 500            \\
 00:11:43   & -10:40:34   & 14.044   & 13.947   & 14.686   & 14.987   & 15.471   & 15.95    & 15.601   & 16.524   & ABU          & 32-35000   & 501         \\
 00:50:40   & -09:25:46   & 12.845    & 12.553   & 13.818   & 14.174   & 14.716   & 15.216   & 15.307   & 15.402   & AAC          & 55000   & 500          \\
 01:11:56   & +15:17:53   & 14.866   & 14.499    & 15.154   & 15.206   & 15.627   & 15.74    & 15.948   & 15.615   & ACD          & 25000   & 501       \\
 01:30:32   & +52:33:50   & 16.679   & 16.402   & 16.072   & 16.067   & 16.18    & 15.931   & 15.965   & 15.998   & ACD          & 37500   & 500          \\
 01:32:33   & +51:57:57   & 15.281   & 15.309    & 14.935   & 14.996   & 15.167   & 15.202   & 15.186   & 15.291   & AAB          & 37500   & 500           \\
 02:20:35   & +17:04:07   & 14.912    & 14.615    & 15.011   & 14.957   & 15.274   & 15.241   & 15.103   & 14.992   & AAC          & 24000   & 500           \\
 02:31:45   & +22:08:30  & 16.749   & 16.597   & 16.545   & 16.392   & 16.646   & 16.319   & 16.277   & 15.914   & ADD          & 23-24000   & 510         \\
 02:34:56   & -06:09:13   & 14.884   & 14.167  & 15.604   & 15.963   & 16.475   & 16.888   & 16.566   & 16.862   & CDU          & 42-55000   & 500        \\
 02:35:48   & +05:35:24   & 14.725   & 14.269   & 15.407   & 15.518   & 15.999   & 16.806   & 16.327   & 15.609   & CCU          & 30000   & 511          \\                    
\hline
%inserts single line
\end{tabular}
\end{table*}

\begin{table*}
\caption{A sample of hot subdwarf candidates with infrared excess in the SED fit. Column labels are like in Table 1. A value of `4' in the first digit of the  {\em Fit flag} column indicates the existence of excess in the red part of the SED. The beginning of the red excess ({\em Excess from} column) is obtained from the first band where the difference between the model and the dereddened observed value is above 20 per cent. The possible spectral type of the companion star is established by comparison with the \O stensen (2006) subdwarf database (see Sec. 3.2). The complete table can be found at \url{http://svo2.cab.inta-csic.es/vocats/hsa/}}
% title of Table
\label{table:2}
% is used to refer this table in the text
\centering
% used for centering table
\begin{tabular}{c c p{0.6cm}  c p{0.5cm} p{0.5cm} c p{0.5cm} p{0.5cm} c c c c c c}
% centered columns (15 columns)
\hline\hline
% inserts double horizontal lines
RA          & DEC         & NUV      & FUV      &  \hspace{0.2cm} u      &\hspace{0.2cm} g      & r      &\hspace{0.2cm} J      & \hspace{0.1cm} H      & K      & 2MASS  & Teff  & Fit  & Excess      & Binary    \\
(J2000)   & (J2000)   &      &   & & & & & & & flag    &   (VOSA) &    flag      & from & class \\
% table heading
\hline
% inserts single horizontal line
 00:14:40 & +08:03:52 & 17.11 & 16.87 & 16.83 & 16.57 & 16.48 & 15.69 & 15.36 & 15.21 & AAB        & 15000 & 410 & z  & GK    \\ 
 01:23:41 & +30:02:32 & 16.02 & 15.62 & 16.33 & 16.35 & 16.64 & 16.42 & 15.84 & 15.29 & BCU        & 18-21000 & 410      & z           & GK    \\ 
 01:24:59 & +47:56:41 & 15.98 & 15.48 & 16.53 & 16.88 & 17.06  & 15.82 & 15.23 & 15.00 & AAA        & 55000 & 400      & i           & FGK    \\ 
 01:33:14 & +48:57:28 & 13.33 & 13.15 & 14.00 & 14.57 & 15.04 & 12.10 & 12.14 & 12.16  & AAA        & 18000 & 410      & B           & F     \\ 
 02:28:23 & +25:35:19 & 14.37 & 13.35 & 13.50 & 14.16 & 13.12  & 13.04 & 13.10 & 13.09 & AAA        & 15000 & 400      & B           &   F    \\ 
 02:41:13 & +21:57:43 & 14.14 & 14.28 & 14.38 & 12.92 & 13.06 & 12.67 & 12.68 & 12.72 & AAA        & 15000 & 400      & B           & F     \\ 
 02:44:14 & +30:07:23 & 15.40 & 14.79 & 15.06 & 14.67 & 14.60 & 13.89 & 13.63 & 13.63 & AAA        & 16000 & 400      & i          & FGK    \\ 
 02:57:48 & +37:15:35 & 15.42 & 14.93 & 16.07  & 16.41 & 16.87 & 16.74  & 16.28 & 15.71 & CDU        & 55000 & 400      & J           & GK    \\ 
 03:18:23 & +41:55:22 & 14.36 & 13.99 & 14.49 & 14.63 & 14.92 & 14.69 & 14.52 & 14.34 & AAA        & 40000 & 410      & J           & GK    \\ 
 03:48:30 & +16:39:46 & 17.89 & 17.58 & 17.41 & 17.22 & 17.07 & 16.15 & 15.57 & 15.79 & ABD        & 21000 & 410      & i           & FGK    \\    
\hline
%inserts single line
\end{tabular}
\end{table*}

\begin{table*}
\caption{A sample of hot subdwarf candidates with bad SED fits.  Column labels are like in Table 1. We include here bad fits with apparent UV excess (`2' as first digit in the {\em Fit flag} column), apparent excess in both UV and IR bands (`3' as first digit) and bad fits of any other sort (`1' as first digit). The complete table can be found at \url{http://svo2.cab.inta-csic.es/vocats/hsa/} }
% title of Table
\label{table:3}
% is used to refer this table in the text
\centering
% used for centering table
\begin{tabular}{c c c  c c c c c c c c c c  }
% centered columns (13 columns)
\hline\hline
% inserts double horizontal lines
RA  & DEC   & NUV   & FUV       & u        & g       & r       & J       & H        & K        & 2MASS   & $T_{\rm eff}$    & Fit       \\
(J2000)   & (J2000)   &      &   & & & & & & & flag    &   (VOSA) &    flag        \\
% table heading
\hline
% inserts single horizontal line
 02:12:44 & +68:07:08 & 19.267 & 18.996 & 19.312 & 19.065 & 18.758  & 15.598 & 14.956 & 15.017 & ABC        & 15000 & 100       \\
  02:25:44 & +72:49:44 & 18.642  & 17.982 & 16.902 & 16.125 & 15.575  & 14.055 & 13.661 & 13.361 & AAA        & 29-37500 & 200         \\
 02:27:18 & +73:36:11 & 17.864  & 17.382 & 16.32  & 15.454 & 14.907  & 13.483 & 12.976 & 12.969 & AAA        & 19000 & 200        \\
 02:51:46 & +75:09:04 & 19.360 & 19.049 & 18.504 & 18.023 & 17.711  & 16.597 & 15.767 & 15.942 & BUU        & 15000 & 200         \\
 03:33:56 & +17:56:36 & 17.699 & 17.156 & 17.975 & 18.011 & 18.26  & 16.703 & 15.656 & 16.914 & BUU        & 18000 & 300       \\
 03:53:07 & +16:48:49 & 15.646 & 15.113 & 15.036 & 14.675 & 14.598  & 13.577 & 13.11  & 13.046 & AAA        & 15000 & 300         \\
 04:07:24 & +14:44:06 & 18.995 & 18.562 & 17.535 & 16.905 & 16.51  & 15.156 & 14.639 & 14.566 & AAA        & 15000 & 300         \\
 04:31:18 & +55:53:08 & 18.696 & 18.598 & 17.365 & 17.054 & 16.889  & 15.899 & 15.531 & 15.277 & ABB        & 37500 & 200        \\
 04:35:42 & +56:24:15 & 19.770 & 19.514  & 18.022 & 17.523 & 17.105  & 15.931 & 15.644 & 15.444 & ABC        & 18000 & 200        \\
 04:38:22 & +19:03:06 & 17.079 & 16.814 & 16.619 & 16.34  & 16.283  & 15.546 & 15.14  & 15.003 & AAB        & 18000 & 310        \\
 04:41:41 & -06:11:29 & 15.389 & 15.132 & 15.735 & 15.54  & 15.614  & 15.061 & 14.833 & 14.867 & AAB        & 15000 & 300       \\ 
\hline
%inserts single line
\end{tabular}
\end{table*}

\begin{table*}
\caption{Hot subdwarf candidates without infrared photometric data. We include here sources with good fit (`5' as first digit in the {\em Fit flag} column), apparent UV excess (`2' as first digit), and  bad fits of any other sort (`1' as first digit).}
% title of Table
\label{table:4}
% is used to refer this table in the text
\centering
% used for centering table
\begin{tabular}{c c c  c c c c c c   }
% centered columns (10 columns)
\hline\hline
% inserts double horizontal lines
RA         & DEC        & NUV  & FUV  & u   & g   & r    & Teff  & Fit      \\
(J2000)   & (J2000)   &      &   & & &      &   (VOSA) &    flag        \\
% table heading
\hline
% inserts single horizontal line
 02:51:03.80 & +75:15:03.4 & 16.7211 & 16.3976 & 17.243 & 17.194 & 17.564 & 42500 & 200  \\
 04:55:29.84 & +24:45:07.6 & 18.8049 & 18.0702 & 18.501 & 18.631 & 18.853  & 42500 & 201  \\
 04:59:12.49 & +60:51:56.5 & 19.0112 & 18.7774 & 19.017 & 18.631 & 18.929  & 37500 & 200  \\
 04:59:39.94 & +59:48:53.5 & 17.5742 & 17.1230 & 17.219 & 17.25  & 17.56   & 42500 & 100  \\
 05:02:51.04 & +13:49:26.9 & 20.4623 & 19.3643 & 19.512 & 19.522 & 19.552  & 42500 & 200  \\
 05:11:25.05 & +15:03:01.0 & 18.9370 & 18.3932 & 17.986 & 18.012 & 18.026  & 40000 & 500  \\
 06:11:51.35 & +34:04:01.5 & 19.3621 & 19.2375 & 19.924 & 20.152 & 20.544  & 42500 & 200  \\
 06:25:53.27 & +34:54:28.2 & 16.8423 & 16.4591 & 17.304 & 17.298 & 17.666  & 42500 & 210  \\
 06:40:51.07 & +26:44:28.0 & 12.9711 & 13.1221 & 13.524 & 11.024 & 11.064  & 15000 & 100  \\
 08:00:03.16 & +07:40:43.3 & 14.6031 & 13.9893 & 15.507 & 15.907 & 16.432  & 55000 & 501  \\
 08:06:08.31 & +10:24:20.1 & 16.4314 & 16.1312 & 17.044 & 17.195 & 17.682   & 28000 & 500  \\
 08:13:32.86 & +05:54:30.1 & 14.8425 & 14.1695 & 15.608 & 16.093 & 16.639  & 55000 & 500  \\
 08:23:15.22 & +00:18:46.0 & 15.6761 & 15.0236 & 16.502 & 17.019 & 17.385 8 & 55000 & 500  \\
 08:28:16.33 & +22:32:26.5 & 14.3126 & 13.6464 & 13.754 & 12.389 & 12.777  & 15000 & 400  \\
 16:07:41.25 & +25:42:20.6 & 16.4320 & 16.1035  & 16.825 & 16.971 & 17.423  & 28000 & 500  \\
 17:37:03.25 & +50:40:41.1 & 15.4195 & 15.2508 & 15.982 & 16.041 & 16.539  & 27000 & 500  \\
 20:14:55.06 & +08:42:13.9 & 18.8628 & 18.5533 & 19.143 & 19.03  & 19.424  & 30000 & 500  \\
 20:46:23.13 & -06:59:26.8 & 16.4524 & 15.9148  & 17.081 & 17.455 & 17.89   & 55000 & 500  \\
 21:08:04.46 & +05:15:28.5 & 16.2921 & 15.8421 & 16.777 & 17.042 & 17.462   & 42500 & 500  \\
 23:28:59.74 & +52:16:24.1 & 18.4366 & 18.0371 & 18.742 & 18.888 & 19.255  & 42500 & 200  \\ 
\hline
%inserts single line
\end{tabular}
\end{table*}

\subsection{Binary sample}
\label{subsec:bin}

An important issue regarding hot subdwarfs is to know the binary fraction of these objects, as some of  the proposed formation channels involve evolution in binary systems \citep{Han2002, Han2003, Clausen12}. 

The three main binary evolution channels, as proposed in these papers,  are the common envelope (CE) ejection channel, the stable Roche lobe overflow (RLOF) channel and the double helium white dwarfs (WDs) merger channel. The CE ejection channel leads to the formation of subdwarfs  in short-period binaries with typical orbital periods between 0.1 and 10 days and very thin hydrogen-rich envelopes. On the other hand, the stable RLOF channel produces  stars with long orbital periods (400 to 1500 days) and with rather thick hydrogen-rich envelopes. The merger channel gives rise to single subdwarfs whose hydrogen-rich envelopes are extremely thin. This channel is believed to explain the formation of helium-rich hot subdwarfs \citep{Zhang2012}. 

Nevertheless, the contribution of the  binary channels to the formation of the different subtypes of hot subdwarf stars is still unclear, with new discoveries challenging the standard binary evolution scenarios (see \citealt{Geier13b} for a recent review).

Regarding sdBs, the binary fraction is estimated around $40$ per cent or higher, depending on the nature of the samples and the method used to detect the stellar companion (see \citealt{Nap2004}, or \citealt{Heber_rev} for a review).  In \cite{Lisker2005} the estimated binary fraction is $32$ per cent, although  this value should be taken as a lower limit as target selection was biased against composite spectrum objects. A recent study on IR excess in known radial velocity (RV) variable sdB binaries can be found in \cite{Kupfer2015}.

The sdO binary fraction is more controversial, with different studies reaching opposite conclusions: \cite{Nap2004} found only one out of 23 RV variable sdO, while \cite{Green2008} and \cite{MUCHFUSS} found a similar distribution of RV variations between sdBs and sdOs.  On the other hand, while \cite{Ulla98} found that 6 out of 14 (43 per cent) new hot sdOs  with IR excesses, \cite{Stro2007} found 8 out of 52 (18 per cent) sdOs with photometric infrared excesses, although again this value should be taken as a lower limit.

 A combination of optical and infrared photometry is commonly used to find late-type companions such as F, G or K types, because the hot subdwarf will shine in the blue, while the companion will have brighter red colours. \cite{Stark2003} found a $40$ per cent of binary systems in a magnitude-limited sample of hot subdwarfs from the Kilkenny catalogue \citep{Kil_cat}, using 2MASS infrared filters and Johnson or Str\"{o}mgren optical photometry. In \cite{Girven2012} they also combine  GALEX ultraviolet photometry with optical and infrared filters to select subdwarf candidates in double systems. They complete the photometric study with a spectroscopic classification, finding a large fraction of composite systems with F, G or K companions. On the other hand, detection of radial velocity variations in sdB stars is used to find close binaries with invisible companions such as white dwarfs (see for instance \citealt{Morales} and \citealt{Copperwheat}).

The photometric data of our sample ranges from the ultraviolet to the far infrared (WISE colours), which enabled us to detect flux excesses from the $B$ magnitude. The companions found with our photometric methodology are expected to be late-type main sequence stars, such as F, G or K.

In our list of 437 hot subdwarf candidates, 20 of them have no infrared photometric data available (see Table \ref{table:4}). For the other 417, we consider as {\sl binary candidates} those with quality SED fit 4 or 3 (IR or both IR and UV excess, Tables \ref{table:2} and \ref{table:3}). There are 189 of them, making a total fraction of $45$ per cent. This fraction could be overestimated, as some of the flux excesses may be apparent, due to inaccurate photometric measures or bad SED fittings. 

The possible spectral type of the companion star was estimated analyzing the excess of hot sds with main sequence companions catalogued in \cite{Roydatabase}. We have selected all subdwarfs classified as sds+F, sds+G or sds+K of this catalogue and identified from which band the excess is detected in VOSA. Our criterion for the companion spectral type  was the following:
\begin{itemize}
\item Excess from $B$,$V$ or $g$ band:  type F (17 objects)
\item Excess from $r$ band: types F, G (6 objects)
\item Excess from $i$ band: types F, G, K (86 objects)
\item Excess from $z$ or $J$ band: types G, K (69 objects)
\item Excess from $H$, $Ks$ or $W1$ band: type K (11 objects)
\end{itemize}

 Hot sds can have companions of other spectral types. Close binaries formed by cool main sequence M-type or substellar objects may alter the measured photometric values. The variation is due to a reflection effect in the light curve caused by the irradiated surface of the much cooler companion. It is estimated that only 1/5 of short period sdBs contain a dM \citep{Roy2013}. In fact, few reflecting sdBs+dM/BD systems are known, and they show typical peak-to-peak photometric amplitudes of $\sim 0.2$ mag or less, diminishing towards blue wavelengths. This effect is not expected to significantly alter our procedure.

Less common are hot subdwarfs in eclipsing binaries, either with dM/BD or WD companions (see \cite{For2010}). In these cases, deep eclipses with $\sim < 0.8$ mag variations occur (although see the extraordinary sdO+dM system in \cite{Derekas15}). A raw estimate of the number of eclipsing sdB+dM systems in our sample leads to a negligible number of photometric disturbances caused by photometry being acquired on eclipse-phase. Of course, we can not rule out that chances of this effect occur, which could explain any of the bad fits encountered in this work. 

A more rigorous check of all possible binary candidates, using a two-component fit,  is presently being addressed, and will be presented in a future work.

%\newpage

\subsection{Colour-colour diagrams}
Colour-colour or colour-magnitude diagrams allow us to separate sources of different nature using photometric colours. This is commonly used to detect ultraviolet or infrared excesses signaling a probable binary nature of the objects under study.

\begin{figure}
\resizebox{\hsize}{!}{\includegraphics{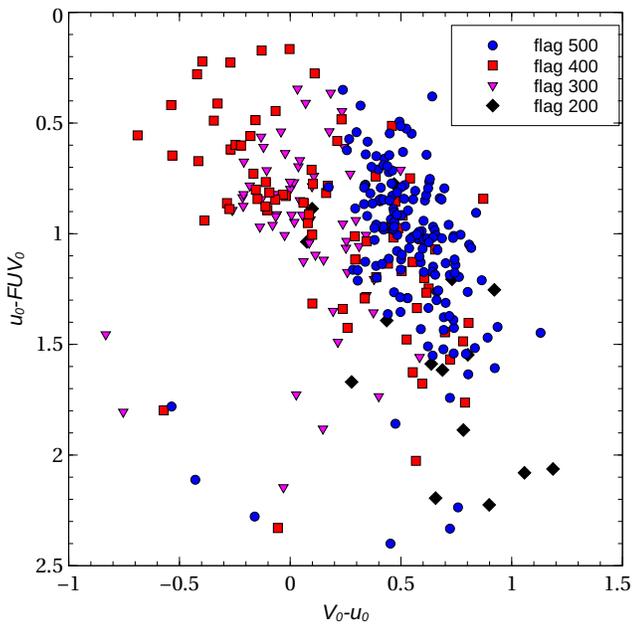}}
\caption{The relative difference in blue colours of our candidates is visible in this plane. The bluer stars lay on the bottom right corner.}
\label{our_colors}
\end{figure}

We  have first plotted our candidates in the colour-colour plane $u_0-FUV_0$ versus $V_0-u_0$, as seen in Fig. \ref{our_colors}. The $0$ subscript stands for derredened magnitudes, with the $E(B-V)$ values taken from \cite{Schlegel}, and the corresponding correction factors computed from the formulae in \cite*{Cardelli}. $FUV$ and $u$ are GALEX and SDSS filters, respectively. As we do not have data of  the $V$ magnitude for all our candidates, we use a  $V$ value computed with the transformation formula given in \cite{Jester2005} between the $ugriz$ and $UBV$ systems:
\begin{equation}
V=g-0.58 (g-r)-0.01 
\label{vsinc}
\end{equation}

The figure shows the relative difference in the blue colours of our candidates. Most of the good fitted objects lay on the right side of the diagram, as would be expected, because they stand for relatively hot single candidates. The bluer stars lay on the bottom right corner of the diagram, where we can see objects labeled with ultraviolet excess (quality fit 200). To discard the UV excess was only an effect of a possible overestimation in the extinction correction, we checked the $A_V$ values of these objects. They turned up to be relatively high ($A_V>1.14$ mag), but not higher than many other objects without UV excess (single or binary candidates).

There are various references in the recent literature studying different samples of hot subdwarfs and proposing alternative ways to separate single from binary stars using colour-colour diagrams. In \cite{Stark2003} they plot $J-K_s$ versus $V-K_s$ and found single stars remain inside a box limited by  $V-K_s \leq +0.2$ and $J-K_s \leq +0.05$, while the composite stars lay outside this rectangle (see fig. 1 in that paper). In \cite{Green2008} they plot $V-J$ versus $J-H$ of a sample of confirmed subdwarfs, finding a gap in the colour diagram, separating single from composite stars (fig.5 of that paper). Finally, in \cite{Girven2012} $FUV-r$ is plotted versus $r-K_s$, showing also separate regions for the single and composite subdwarfs.

\begin{figure}
\resizebox{\hsize}{!}{\includegraphics{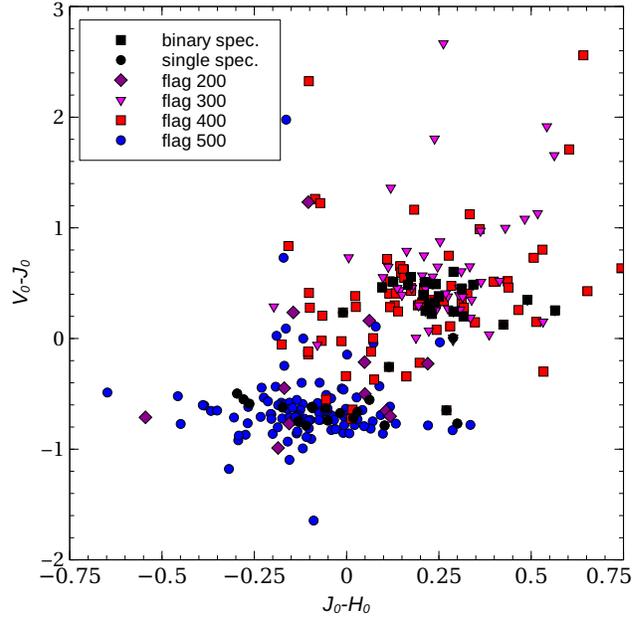}}
\caption{An analogous of the gap in Green et al. (2008) is presented here. Single subdwarf candidates (flag 500) concentrate in the bottom-left of the diagram, while composite candidates (flag 400 and 300) are in the upper right side. We also represent the binaries and single objects spectroscopically identified using SDSS spectra (see Sect. 4).}
\label{green}
\end{figure}

Similarly to \cite{Green2008}, to show the single or binary nature of our candidates we plotted $V_0-J_0$ versus $J_0-H_0$, the $V$ magnitude computed as in equation (\ref{vsinc}) and $J$ and $H$ from 2MASS.  The result is shown in Fig. \ref{green}. 

Not all the candidates are represented there. Many of our sources are quite faint in the infrared, and thus do not have very good 2MASS flags: 173 sources ($41$ per cent) have at least an $U$ ({\it upper limit}) in one of the 2MASS filters; while other 46 sources ($11$ per cent) have at least one filter with a $D$ or $E$ flag\footnote{\url{http://www.ipac.caltech.edu/2mass/releases/allsky/doc/sec1_6b.html} gives detailed explanations of their quality flags.}. To avoid big errors, we  restricted our plot to sources with $err(J)+err(H)<0.3$. For the rest of the objects,  we substituted 2MASS data with UKIDSS, if available.

In Fig. \ref{green}, we can see that single candidates tend to concentrate around  $V_0-J_0 \simeq -0.6$, and binary candidates, although more spread, around  $V_0-J_0 \simeq 0.25$.  In spite of the source spreading, the gap between composite and single systems is still visible. The source spreading  may be due to a variety of factors. We first noted that the majority of binary candidates spread around the diagram are those where the excess begins at $B$ or $r$ magnitudes, causing the optical region of the SED to be untypical. Other uncertainties in the SDSS photometry, due to either faint magnitudes  or bright values near the saturation limits, could cause the star to be misplaced. Finally, we can not discard other phenomena, like the presence of planetary nebulae.

\section{Spectral classification}

A first classification scheme for hot subdwarfs was proposed in \cite{PGcat}, where an eight class system was defined. In \cite{Drilling13}  an evolved MK-like system is developed. In this system hot subdwarfs are divided in four sequences: {\sl He-weak}, {\sl He-normal}, {\sl He-strong} and {\sl He-strong C}, with carbon or other metallic lines. The differences in helium content of each object are also quantified measuring ratios of hydrogen to helium line depths. In each of the four helium sequences, the spectral subclasses would range from sdO1 to sdB9, as in the MK system, line depths and ratios varying smoothly within each sequence (see figs. 1 to 4 in that paper). Stars former classified within the somehow {\it ad-hoc} sdOB subclass \citep{Moehler1990}  are placed naturally in this scheme, in the transition between late sdO and early sdB subclasses. In the present work we will appply the \cite{Drilling13} system to classify the candidates with spectrum.

Only 68 stars  of our list of subdwarf candidates ($16$ per cent) had SDSS spectrum.  We begun with a visual inspection of each object's whole spectrum. One spectrum was too noisy to allow identification. The rest were identified as one white dwarf, one probable cataclysmic variable and 65 subdwarfs: 5 sdOs ($8$ per cent), 25 sdOBs ($38$ per cent) and 35 proper sdBs ($54$ per cent).  Note the effectiveness of our selection procedure improves to $95.6$ per cent within this subset.

We also inspected the presence of characteristic lines of cooler main sequence stars, to identify binary candidates. In particular, we looked for the  Mg\,{\sc i} triplet (5172, 5183, 5167\AA), the G band (4300\AA), and the  Ca\,{\sc ii} K line (3933\AA). The Na\,{\sc i} doublet (5889-5895\AA) can also be an indicator, although it may be overlapped with a near He\,{\sc ii} line. We found 23 probable binary systems: 22 binary sdBs (including sdOBs) and 1 binary sdOs. The binary fraction for sdBs obtained by visual inspection of the spectrum was of $37$ per cent, somehow lower than the photometric fraction. Besides the possible overestimation of the photometric fraction, as argued above, the relatively low signal-to-noise ratio of some SDSS spectra may be obscuring the binary nature of some candidates, hiding the cold companion lines. We point out that all the spectrum-detected binary systems are also binary candidates from the photometric excess point of view.

\begin{figure}
\resizebox{\hsize}{!}{\includegraphics{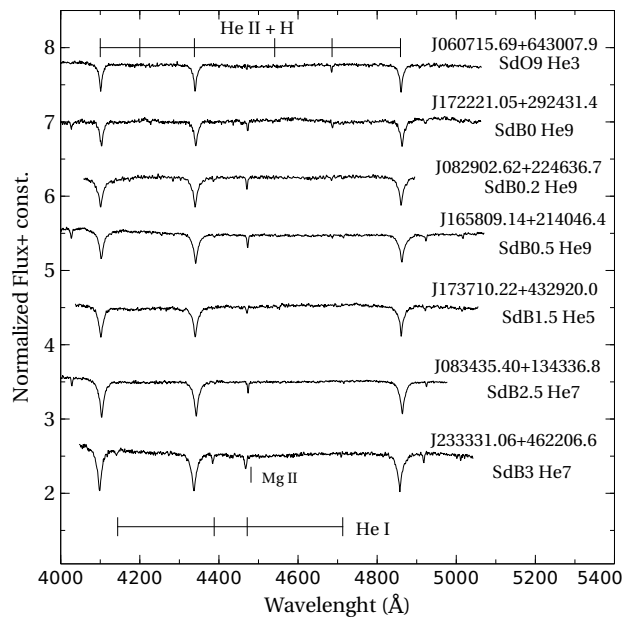}}
\caption{Subdwarfs in the {\sl He-normal} spectral sequence, following Drilling et al. (2013).  Star identifications are taken from SDSS.}
\label{normal}
\end{figure}

\begin{figure}
\resizebox{\hsize}{!}{\includegraphics{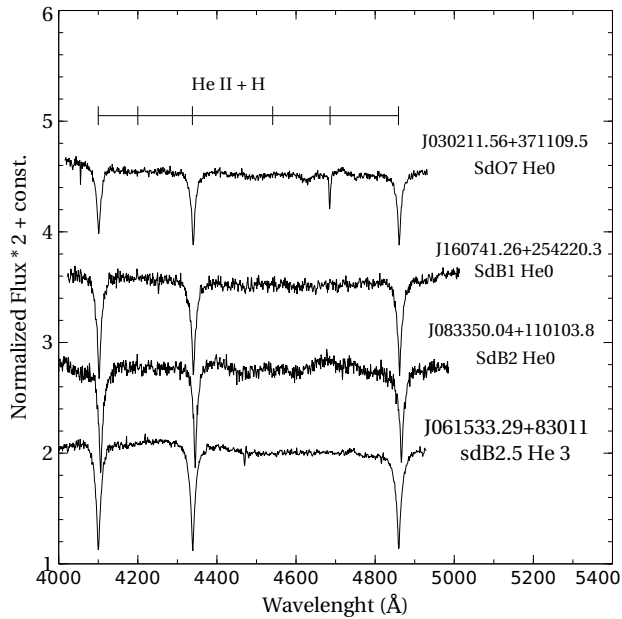}}
\caption{Subdwarfs in the {\sl He-weak} spectral sequence, following Drilling et al. (2013).  Star identifications are taken from SDSS.}
\label{weak}
\end{figure}

To classify the subdwarfs within the \cite{Drilling13} system we cut and normalized the spectra between 4000 and 5000 $\AA$ and then compared them with the standard stars defining the system. The complete classification of our candidates can be found in the electronic tables. A sample is shown in Figures \ref{normal}-\ref{weak}. Sumarizing our results, we have found:
\begin{itemize}
\item 1 star ($1.5$ per cent) belonging to the {\sl He-strong C} sequence,
\item 2 stars ($3.1$ per cent) in the {\sl He-strong } sequence,
\item 48 stars ($73.9$ per cent) in the {\sl He-normal} sequence and
\item 14 stars ($21.5$ per cent) in the {\sl He-weak} sequence.
\end{itemize}
Note that the signal-to-noise ratio of some spectra might be masking weak metallic carbon or nitrogen lines, affecting the {\sl He-strong C} and {\sl He-strong} relative abundances. 

Calibrations made in \cite{Drilling13} demonstrate that {\sl He-weak} stars have subsolar helium abundances, {\sl He-normal} stars more nearly solar abundances and {\sl He-strong} objects high helium abundances. The problem of helium abundances in hot subdwarfs has been addressed in \cite{Toole2008} and \cite{Geier13}. In \cite{Edelmann2003} a correlation between effective temperature and helium abundance in sdB stars was discovered, showing two sequences with approximately the same trend with increasing $T_{\rm eff}$ (fig. 5 in that paper).  The ratio of objects in the lower sequence to those in the upper one is argued to be between 1:10 and 4:10 in  \cite{Toole2008}. Our ratio 14:48 of weak to normal helium stars is consistent with these margins.

\cite{Drilling13} suggest their spectral sequences are temperature sequences, and find linear trends plotting the effective temperature against the spectral class (fig. 10 and 11 in that paper). For a check, we plotted our good SED fitted candidates in the $T_{\rm eff}$-spectral class plane, together with \cite{Drilling13} linear regressions. As a result, and although our temperatures appear in general subestimated, we found our candidates from classes sdO9 to sdB3 to roughly follow Drilling et al.'s  trend lines. On the contrary, the hotter sdO3-sdO7 classes do not seem to follow this behavior. Such a result is not surprising, due to the effective temperature upper limit of the TLUSTY models ($55000K$).

%\begin{figure}\resizebox{\hsize}{!}{\includegraphics{figure6.eps}}\caption{Subdwarfs in the {\sl He-strong C} spectral sequence, following Drilling et al. (2013). Star identifications are taken from SDSS.}\label{strongC}\end{figure}


\section{Conclusions}

In this work we have  extended the selection procedure developed in \cite{Oreiro2011} to identify hot subdwarfs, taking advantage of Virtual Observatory tools. The selection procedure includes photometric and proper motions filters and an effective temperature cutoff. We have identified 437 new subdwarf candidates from a 11663 $deg^2$ sky region, limited by the SDSS DR7 image coverage. We expect an effectiveness of at least $80$ per cent, although the subsample of objects with SDSS spectra reached a subdwarf identification above $95$ per cent, proving the accuracy of our selection filters. 

From our spectral energy distribution analysis, we have estimated a photometric binary fraction of $45$ per cent, while identification of cool star metallic lines in the spectra yields a $37$ per cent of binaries among this subsample. Both numbers, although rough estimates, are in agreement with previous studies. Our method mainly selects binaries with late-type main sequence stellar companions, like F, G or K. 

The colour-colour diagrams of Fig. \ref{our_colors}  and \ref{green} show the difference in, respectively, the blue and red colours of our subdwarf candidates. Objects in the lower right corner of Fig. \ref{our_colors} show a clear ultraviolet excess (labeled with SED fitting flag 200 -- see text for details). In Fig. \ref{green} we see a clear clustering of the single candidates around  $V_0-J_0 \simeq -0.6$, and the colour gap between single and binary subdwarfs discovered in \cite{Green2008} is reproduced.

We have also performed a detailed spectral classification of the 68 candidates with SDSS spectra, following a recent MK-like system for hot subdwarfs developed by \cite{Drilling13}. We found our candidates perfectly suit in one of the four helium sequences proposed in that paper.

Much work remains to be done after the selection procedure developed here. Regarding the list of hot subdwarf candidates, a deeper analysis of the binary sample is presently under study. A two-body SED fitting will yield temperature values of the cold companion, which will aid to estimate its spectral class and its distance. As argued in \cite{Clausen12}, composite systems of the type sdB + early F could be crucial in determining the binary formation channel of hot subdwarfs, depending on the periods measured for these systems. Recently, in \cite{Barlow13} and \cite{Vos13},  long period sdB+F/G systems have been reported. It would thus be interesting to know if our sample contains this type of binaries, and to make a follow-up study of them, determining their orbital parameters.

Both the twenty candidates without available infrared photometry and the seventeen objects with apparent ultraviolet excess are very interesting from our point of view, as good candidates for very hot objects. The hotter sdOs are measured to have up to $100000 K$ effective temperatures \citep{Stro2007}. These objects are scarce between subdwarfs,  and some of them have been proved to have planetary nebulae \citep{Alba13, Alba15}, a signal of a probable post AGB origin \citep{Heber91}. Photometric and spectroscopic accurate data would be needed to reach further conclusions about the origin of these stars. 

Finally, a detailed line spectral analysis of the hot subdwarf candidates, to be performed using advanced/accurate NLTE atmospheric models, would yield more reliable values for the star effective temperatures, helium abundances and surface gravities. The position of our candidates in the $T_{\rm eff}-\log g$ plane would contribute to discriminate between the different origins and evolution paths proposed for hot subdwarfs.

Regarding the search of new hot subdwarf candidates, different approaches could be used. One of those would imply applying our selection procedure to new releases of some of the surveys already considered (e.g. SDSS, GALEX) or using new catalogues, both in the optical (e.g. Pan-Starrs, or J-PAS in the near future) and in the infrared (UKIDSS, VISTA).
 
Another possibility would be employing other catalogues containing astrometric information. At this moment, our routine discards any source without proper motion data in the SuperCosmos survey. This fact could be modified making the routine look for proper motion information in other catalogues, like UCAC4 or PPMXL \citep*{PPMXL}. 

It would also be interesting to explore the possibilities of the data that GAIA\footnote{\url{http://sci.esa.int/gaia/}} would provide regarding this point. GAIA would measure  distances with great accuracy, giving us information of star luminosities and their position in the HR diagram. This would yield estimations of both star masses and ages, an important information to track the evolutionary paths of hot subdwarfs. 

Attention must be paid as well to other forthcoming missions such as CHEOPS \citep{Fortier14} or PLATO \citep{PLATO}, as they are expected to have an impact on ultra-high precision photometry and stellar astroseismology for brigth targets, covering large fractions of the sky (up to 50 per cent in the case of PLATO), and widening, then, the possible detection of new pulsating hot sds.

\section*{Acknowledgements}

We thank A. Aller for helping us improving the spectra classification. This publication makes use of VOSA, developed under the Spanish Virtual Observatory project supported from the Spanish MICINN through grant AyA2011-24052. It has also been partially supported by grant INCITE09312191PR (which includes FEDER funds), given by the Xunta de Galicia, and by grant 12VI20, given by the Universidade de Vigo. This research has made use of the SIMBAD database, the Vizier catalogue access tool (originally published in \citealt*{Vizier}) and the {\sl Aladin sky atlas}, all operated at CDS, Strasbourg, France.  

\bibliographystyle{mn2e} 

\bibliography{references}

\begin{thebibliography}{73}
\expandafter\ifx\csname natexlab\endcsname\relax\def\natexlab#1{#1}\fi

\bibitem[{{Abazajian} {et~al}\mbox{.}(2009){Abazajian}, {Adelman-McCarthy},
  {Ag{\"u}eros}, {Allam}, {Allende Prieto}, {An}, {Anderson}, {Anderson},
  {Annis}, {Bahcall}  \& et~al.}]{DR7}
{Abazajian} K.~N. {et~al.}, 2009, \apjs, 182, 543

\bibitem[{{Aller} {et~al}\mbox{.}(2015){Aller}, {Miranda}, {Olgu{\'{\i}}n},
  {V{\'a}zquez}, {Guill{\'e}n}, {Oreiro}, {Ulla}  \& {Solano}}]{Alba15}
{Aller} A., {Miranda} L.~F., {Olgu{\'{\i}}n} L., {V{\'a}zquez} R.,
  {Guill{\'e}n} P.~F., {Oreiro} R., {Ulla} A., {Solano} E., 2015, \mnras, 446,
  317

\bibitem[{{Aller} {et~al}\mbox{.}(2013){Aller}, {Miranda}, {Ulla},
  {V{\'a}zquez}, {Guill{\'e}n}, {Olgu{\'{\i}}n}, {Rodr{\'{\i}}guez-L{\'o}pez},
  {Thejll}, {Oreiro}, {Manteiga}  \& {P{\'e}rez}}]{Alba13}
{Aller} A. {et~al.}, 2013, \aap, 552, A25

\bibitem[{{Arce} \& {Goodman}(1999)}]{Arce99}
{Arce} H.~G., {Goodman} A.~A., 1999, \apjl, 512, L135

\bibitem[{{Barlow} {et~al}\mbox{.}(2013){Barlow}, {Liss}, {Wade}  \&
  {Green}}]{Barlow13}
{Barlow} B.~N., {Liss} S.~E., {Wade} R.~A., {Green} E.~M., 2013, \apj, 771, 23

\bibitem[{{Bayo} {et~al}\mbox{.}(2008){Bayo}, {Rodrigo}, {Barrado y
  Navascu{\'e}s}, {Solano}, {Guti{\'e}rrez}, {Morales-Calder{\'o}n}  \&
  {Allard}}]{VOSA}
{Bayo} A., {Rodrigo} C., {Barrado y Navascu{\'e}s} D., {Solano} E.,
  {Guti{\'e}rrez} R., {Morales-Calder{\'o}n} M., {Allard} F., 2008, \aap, 492,
  277

\bibitem[{{Bianchi} \& {GALEX Team}(2000)}]{Galex}
{Bianchi} L., {GALEX Team}, 2000, \memsai, 71, 1123

\bibitem[{{Brown} {et~al}\mbox{.}(1997){Brown}, {Ferguson}, {Davidsen}  \&
  {Dorman}}]{Brown1997}
{Brown} T.~M., {Ferguson} H.~C., {Davidsen} A.~F., {Dorman} B., 1997, \apj,
  482, 685

\bibitem[{{Camarota} \& {Holberg}(2014)}]{Camarota2014}
{Camarota} L., {Holberg} J.~B., 2014, \mnras, 438, 3111

\bibitem[{{Cardelli} {et~al}\mbox{.}(1989){Cardelli}, {Clayton}  \&
  {Mathis}}]{Cardelli}
{Cardelli} J.~A., {Clayton} G.~C., {Mathis} J.~S., 1989, \apj, 345, 245

\bibitem[{{Castelli} {et~al}\mbox{.}(1997){Castelli}, {Gratton}  \&
  {Kurucz}}]{Castelli97}
{Castelli} F., {Gratton} R.~G., {Kurucz} R.~L., 1997, \aap, 318, 841

\bibitem[{{Clausen} {et~al}\mbox{.}(2012){Clausen}, {Wade}, {Kopparapu}  \&
  {O'Shaughnessy}}]{Clausen12}
{Clausen} D., {Wade} R.~A., {Kopparapu} R.~K., {O'Shaughnessy} R., 2012, \apj,
  746, 186

\bibitem[{{Copperwheat} {et~al}\mbox{.}(2011){Copperwheat}, {Morales-Rueda},
  {Marsh}, {Maxted}  \& {Heber}}]{Copperwheat}
{Copperwheat} C.~M., {Morales-Rueda} L., {Marsh} T.~R., {Maxted} P.~F.~L.,
  {Heber} U., 2011, \mnras, 415, 1381

\bibitem[{{D'Cruz} {et~al}\mbox{.}(1996){D'Cruz}, {Rood}, {O'Connell}, {Dorman}
   \& {Dickens}}]{Cruz96}
{D'Cruz} N., {Rood} R., {O'Connell} R., {Dorman} B., {Dickens} R., 1996, in
  Bulletin of the American Astronomical Society, Vol.~28, American Astronomical
  Society Meeting Abstracts, p. 1390

\bibitem[{{Derekas} {et~al}\mbox{.}(2015){Derekas}, {Nemeth}, {Southworth},
  {Borkovits}, {Sarneczky}, {Pal}, {Csak}, {Garcia-Alvarez}, {Maxted}, {Kiss},
  {Vida}, {Szabo}  \& {Kriskovics}}]{Derekas15}
{Derekas} A. {et~al.}, 2015, preprint (arXiv:1505.06487)

\bibitem[{{Downes} {et~al}\mbox{.}(2001){Downes}, {Webbink}, {Shara}, {Ritter},
  {Kolb}  \& {Duerbeck}}]{Downes}
{Downes} R.~A., {Webbink} R.~F., {Shara} M.~M., {Ritter} H., {Kolb} U.,
  {Duerbeck} H.~W., 2001, \pasp, 113, 764

\bibitem[{{Drilling} {et~al}\mbox{.}(2013){Drilling}, {Jeffery}, {Heber},
  {Moehler}  \& {Napiwotzki}}]{Drilling13}
{Drilling} J.~S., {Jeffery} C.~S., {Heber} U., {Moehler} S., {Napiwotzki} R.,
  2013, \aap, 551, A31

\bibitem[{{Edelmann} {et~al}\mbox{.}(2003){Edelmann}, {Heber}, {Hagen},
  {Lemke}, {Dreizler}, {Napiwotzki}  \& {Engels}}]{Edelmann2003}
{Edelmann} H., {Heber} U., {Hagen} H.-J., {Lemke} M., {Dreizler} S.,
  {Napiwotzki} R., {Engels} D., 2003, \aap, 400, 939

\bibitem[{{Fitzpatrick}(1999)}]{Fitz99}
{Fitzpatrick} E.~L., 1999, \pasp, 111, 63

\bibitem[{{For} {et~al}\mbox{.}(2010){For}, {Green}, {Fontaine}, {Drechsel},
  {Shaw}, {Dittmann}, {Fay}, {Francoeur}, {Laird}, {Moriyama}, {Morris},
  {Rodr{\'{\i}}guez-L{\'o}pez}, {Sierchio}, {Story}, {Strom}, {Wang}, {Adams},
  {Bolin}, {Eskew}  \& {Chayer}}]{For2010}
{For} B.-Q. {et~al.}, 2010, \apj, 708, 253

\bibitem[{{Fortier} {et~al}\mbox{.}(2014){Fortier}, {Beck}, {Benz}, {Broeg},
  {Cessa}, {Ehrenreich}  \& {Thomas}}]{Fortier14}
{Fortier} A., {Beck} T., {Benz} W., {Broeg} C., {Cessa} V., {Ehrenreich} D.,
  {Thomas} N., 2014, in Society of Photo-Optical Instrumentation Engineers
  (SPIE) Conference Series, Vol. 9143, Society of Photo-Optical Instrumentation
  Engineers (SPIE) Conference Series, p.~2

\bibitem[{{Geier} {et~al}\mbox{.}(2011){Geier}, {Hirsch}, {Tillich}, {Maxted},
  {Bentley}, {{\O}stensen}, {Heber}, {G{\"a}nsicke}, {Marsh}, {Napiwotzki},
  {Barlow}  \& {O'Toole}}]{MUCHFUSS}
{Geier} S. {et~al.}, 2011, \aap, 530, A28

\bibitem[{{Geier}(2013)}]{Geier13b}
{Geier} S., 2013, in European Physical Journal Web of Conferences, Vol.~43, p.
  4001

\bibitem[{{Geier} {et~al}\mbox{.}(2013){Geier}, {Heber}, {Edelmann},
  {Morales-Rueda}, {Kilkenny}, {O'Donoghue}, {Marsh}  \&
  {Copperwheat}}]{Geier13}
{Geier} S., {Heber} U., {Edelmann} H., {Morales-Rueda} L., {Kilkenny} D.,
  {O'Donoghue} D., {Marsh} T.~R., {Copperwheat} C., 2013, \aap, 557, A122



\bibitem[{{Geier} {et~al}\mbox{.}(2015){Geier}, {Kupfer}, {Heber},
  {Schaffenroth}, {Barlow}, {{\O}stensen}, {O'Toole}, {Ziegerer}, {Heuser},
  {Maxted}, {G{\"a}nsicke}, {Marsh}, {Napiwotzki}, {Br{\"u}nner}, {Schindewolf}
   \& {Niederhofer}}]{Geier15_RVcat}
{Geier} S. {et~al.}, 2015, \aap, 577, A26

\bibitem[{{Gentile Fusillo} {et~al}\mbox{.}(2015){Gentile Fusillo},
  {G{\"a}nsicke}  \& {Greiss}}]{Gentile2015}
{Gentile Fusillo} N.~P., {G{\"a}nsicke} B.~T., {Greiss} S., 2015, \mnras, 448,
  2260

\bibitem[{{Girven} {et~al}\mbox{.}(2012){Girven}, {Steeghs}, {Heber},
  {G{\"a}nsicke}, {Marsh}, {Breedt}, {Copperwheat}, {Pyrzas}  \&
  {Longa-Pe{\~n}a}}]{Girven2012}
{Girven} J. {et~al.}, 2012, \mnras, 425, 1013

\bibitem[{{Green} {et~al}\mbox{.}(2008){Green}, {Fontaine}, {Hyde}, {For}  \&
  {Chayer}}]{Green2008}
{Green} E.~M., {Fontaine} G., {Hyde} E.~A., {For} B.-Q., {Chayer} P., 2008, in
  Astronomical Society of the Pacific Conference Series, Vol. 392, Hot Subdwarf
  Stars and Related Objects, {Heber} U., {Jeffery} C.~S., {Napiwotzki} R.,
  eds., p.~75

\bibitem[{{Green} {et~al}\mbox{.}(1986){Green}, {Schmidt}  \&
  {Liebert}}]{PGcat}
{Green} R.~F., {Schmidt} M., {Liebert} J., 1986, \apjs, 61, 305

\bibitem[{{Hambly} {et~al}\mbox{.}(2001){Hambly}, {MacGillivray}, {Read},
  {Tritton}, {Thomson}, {Kelly}, {Morgan}, {Smith}, {Driver}, {Williamson},
  {Parker}, {Hawkins}, {Williams}  \& {Lawrence}}]{SuperCosmos}
{Hambly} N.~C. {et~al.}, 2001, \mnras, 326, 1279

\bibitem[{{Han} {et~al}\mbox{.}(2003){Han}, {Podsiadlowski}, {Maxted}  \&
  {Marsh}}]{Han2003}
{Han} Z., {Podsiadlowski} P., {Maxted} P.~F.~L., {Marsh} T.~R., 2003, \mnras,
  341, 669

\bibitem[{{Han} {et~al}\mbox{.}(2002){Han}, {Podsiadlowski}, {Maxted}, {Marsh}
  \& {Ivanova}}]{Han2002}
{Han} Z., {Podsiadlowski} P., {Maxted} P.~F.~L., {Marsh} T.~R., {Ivanova} N.,
  2002, \mnras, 336, 449

\bibitem[{{Heber}(1991)}]{Heber91}
{Heber} U., 1991, in IAU Symposium, Vol. 145, Evolution of Stars: the
  Photospheric Abundance Connection, {Michaud} G., {Tutukov} A.~V., eds., p.
  363

\bibitem[{{Heber} {et~al}\mbox{.}(2000){Heber}, {Reid}  \&
  {Werner}}]{Heber2000}
{Heber} U., {Reid} I.~N., {Werner} K., 2000, \aap, 363, 198

\bibitem[{{Heber}(2009)}]{Heber_rev}
{Heber} U., 2009, \araa, 47, 211



\bibitem[{{H{\o}g} {et~al}\mbox{.}(2000){H{\o}g}, {Fabricius}, {Makarov},
  {Bastian}, {Schwekendiek}, {Wicenec}, {Urban}, {Corbin}  \& {Wycoff}}]{Tycho}
{H{\o}g} E. {et~al.}, 2000, \aap, 357, 367

\bibitem[{{Hubeny} \& {Lanz}(1995)}]{Hubeny1995}
{Hubeny} I., {Lanz} T., 1995, \apj, 439, 875

\bibitem[{{Humason} \& {Zwicky}(1947)}]{Humason47}
{Humason} M.~L., {Zwicky} F., 1947, \apj, 105, 85

\bibitem[{{Jester} {et~al}\mbox{.}(2005){Jester}, {Schneider}, {Richards},
  {Green}, {Schmidt}, {Hall}, {Strauss}, {Vanden Berk}, {Stoughton}, {Gunn},
  {Brinkmann}, {Kent}, {Smith}, {Tucker}  \& {Yanny}}]{Jester2005}
{Jester} S. {et~al.}, 2005, \aj, 130, 873

\bibitem[{{Kawka} {et~al}\mbox{.}(2015){Kawka}, {Vennes}, {O'Toole},
  {N{\'e}meth}, {Burton}, {Kotze}  \& {Buckley}}]{Kawka2015}
{Kawka} A., {Vennes} S., {O'Toole} S., {N{\'e}meth} P., {Burton} D., {Kotze}
  E., {Buckley} D.~A.~H., 2015, \mnras, 450, 3514

\bibitem[{{Kepler} {et~al}\mbox{.}(2016){Kepler}, {Pelisoli}, {Koester},
  {Ourique}, {Romero}, {Reindl}, {Kleinman}, {Eisenstein}, {Valois}  \&
  {Amaral}}]{Kepler15}
{Kepler} S.~O. {et~al.}, 2016, \mnras, 455, 3413

\bibitem[{{Kilkenny}(2002)}]{pulsating}
{Kilkenny} D., 2002, in Astronomical Society of the Pacific Conference Series,
  Vol. 259, IAU Colloq. 185: Radial and Nonradial Pulsationsn as Probes of
  Stellar Physics, {Aerts} C., {Bedding} T.~R., {Christensen-Dalsgaard} J.,
  eds., p. 356

\bibitem[{{Kilkenny} {et~al}\mbox{.}(1988){Kilkenny}, {Heber}  \&
  {Drilling}}]{Kil_cat}
{Kilkenny} D., {Heber} U., {Drilling} J.~S., 1988, South African Astronomical
  Observatory Circular, 12, 1

\bibitem[{{Kleinman} {et~al}\mbox{.}(2013){Kleinman}, {Kepler}, {Koester},
  {Pelisoli}, {Pe{\c c}anha}, {Nitta}, {Costa}, {Krzesinski}, {Dufour},
  {Lachapelle}, {Bergeron}, {Yip}, {Harris}, {Eisenstein}, {Althaus}  \&
  {C{\'o}rsico}}]{Kleinman}
{Kleinman} S.~J. {et~al.}, 2013, \apjs, 204, 5

\bibitem[{{Kupfer} {et~al}\mbox{.}(2015){Kupfer}, {Geier}, {Heber},
  {{\O}stensen}, {Barlow}, {Maxted}, {Heuser}, {Schaffenroth}  \&
  {G{\"a}nsicke}}]{Kupfer2015}
{Kupfer} T. {et~al.}, 2015, \aap, 576, A44

\bibitem[{{Lanz} \& {Hubeny}(2003)}]{Lanz2003}
{Lanz} T., {Hubeny} I., 2003, \apjs, 146, 417

\bibitem[{{Lanz} \& {Hubeny}(2007)}]{Lanz2007}
{Lanz} T., {Hubeny} I., 2007, \apjs, 169, 83

\bibitem[{{Lawrence} {et~al}\mbox{.}(2007){Lawrence}, {Warren}, {Almaini},
  {Edge}, {Hambly}, {Jameson}, {Lucas}, {Casali}, {Adamson}, {Dye}, {Emerson},
  {Foucaud}, {Hewett}, {Hirst}, {Hodgkin}, {Irwin}, {Lodieu}, {McMahon},
  {Simpson}, {Smail}, {Mortlock}  \& {Folger}}]{UKIDSS}
{Lawrence} A. {et~al.}, 2007, \mnras, 379, 1599

\bibitem[{{Lisker} {et~al}\mbox{.}(2005){Lisker}, {Heber}, {Napiwotzki},
  {Christlieb}, {Han}, {Homeier}  \& {Reimers}}]{Lisker2005}
{Lisker} T., {Heber} U., {Napiwotzki} R., {Christlieb} N., {Han} Z., {Homeier}
  D., {Reimers} D., 2005, \aap, 430, 223

\bibitem[{{McCook} \& {Sion}(1999)}]{Cook}
{McCook} G.~P., {Sion} E.~M., 1999, \apjs, 121, 1

\bibitem[{{Mengel} {et~al}\mbox{.}(1976){Mengel}, {Norris}  \&
  {Gross}}]{Mengel96}
{Mengel} J.~G., {Norris} J., {Gross} P.~G., 1976, \apj, 204, 488

\bibitem[{{Moehler} {et~al}\mbox{.}(1990){Moehler}, {Richtler}, {de Boer},
  {Dettmar}  \& {Heber}}]{Moehler1990}
{Moehler} S., {Richtler} T., {de Boer} K.~S., {Dettmar} R.~J., {Heber} U.,
  1990, \aaps, 86, 53

\bibitem[{{Morales-Rueda} {et~al}\mbox{.}(2003){Morales-Rueda}, {Maxted},
  {Marsh}, {North}  \& {Heber}}]{Morales}
{Morales-Rueda} L., {Maxted} P.~F.~L., {Marsh} T.~R., {North} R.~C., {Heber}
  U., 2003, \mnras, 338, 752

\bibitem[{{Napiwotzki} {et~al}\mbox{.}(2004){Napiwotzki}, {Karl}, {Lisker},
  {Heber}, {Christlieb}, {Reimers}, {Nelemans}  \& {Homeier}}]{Nap2004}
{Napiwotzki} R., {Karl} C.~A., {Lisker} T., {Heber} U., {Christlieb} N.,
  {Reimers} D., {Nelemans} G., {Homeier} D., 2004, \apss, 291, 321

\bibitem[{{N{\'e}meth} {et~al}\mbox{.}(2012){N{\'e}meth}, {Kawka}  \&
  {Vennes}}]{Nemeth2012}
{N{\'e}meth} P., {Kawka} A., {Vennes} S., 2012, \mnras, 427, 2180

\bibitem[{{Ochsenbein} {et~al}\mbox{.}(2000){Ochsenbein}, {Bauer}  \&
  {Marcout}}]{Vizier}
{Ochsenbein} F., {Bauer} P., {Marcout} J., 2000, \aaps, 143, 23

\bibitem[{{Oreiro} {et~al}\mbox{.}(2011){Oreiro}, {Rodr{\'{\i}}guez-L{\'o}pez},
  {Solano}, {Ulla}, {{\O}stensen}  \& {Garc{\'{\i}}a-Torres}}]{Oreiro2011}
{Oreiro} R., {Rodr{\'{\i}}guez-L{\'o}pez} C., {Solano} E., {Ulla} A.,
  {{\O}stensen} R., {Garc{\'{\i}}a-Torres} M., 2011, \aap, 530, A2

\bibitem[{{{\O}stensen}(2006)}]{Roydatabase}
{{\O}stensen} R.~H., 2006, Baltic Astronomy, 15, 85

\bibitem[{{{\O}stensen} {et~al}\mbox{.}(2013){{\O}stensen}, {Geier},
  {Schaffenroth}, {Telting}, {Bloemen}, {N{\'e}meth}, {Beck}, {Lombaert},
  {P{\'a}pics}, {Tillich}, {Ziegerer}, {Fox Machado}, {Littlefair}, {Dhillon},
  {Aerts}, {Heber}, {Maxted}, {G{\"a}nsicke}  \& {Marsh}}]{Roy2013}
{{\O}stensen} R.~H. {et~al.}, 2013, \aap, 559, A35

\bibitem[{{O'Toole}(2008)}]{Toole2008}
{O'Toole} S.~J., 2008, in Astronomical Society of the Pacific Conference
  Series, Vol. 392, Hot Subdwarf Stars and Related Objects, {Heber} U.,
  {Jeffery} C.~S., {Napiwotzki} R., eds., p.~67

\bibitem[{{Rauer} {et~al}\mbox{.}(2014){Rauer}, {Catala}, {Aerts},
  {Appourchaux}, {Benz}, {Brandeker}, {Christensen-Dalsgaard}, {Deleuil},
  {Gizon}, {Goupil}, {G{\"u}del}, {Janot-Pacheco}, {Mas-Hesse}, {Pagano},
  {Piotto}, {Pollacco}, {Santos}, {Smith}, {Su{\'a}rez}, {Szab{\'o}}, {Udry},
  {Adibekyan}, {Alibert}, {Almenara}, {Amaro-Seoane}, {Eiff}, {Asplund},
  {Antonello}, {Barnes}  \& {Baudin}}]{PLATO}
{Rauer} H. {et~al.}, 2014, Experimental Astronomy, 38, 249

\bibitem[{{Reed}(2005)}]{Reed}
{Reed} C., 2005, VizieR Online Data Catalog, 5125, 0

\bibitem[{{Roeser} {et~al}\mbox{.}(2010){Roeser}, {Demleitner}  \&
  {Schilbach}}]{PPMXL}
{Roeser} S., {Demleitner} M., {Schilbach} E., 2010, \aj, 139, 2440

\bibitem[{{Schlegel} {et~al}\mbox{.}(1998){Schlegel}, {Finkbeiner}  \&
  {Davis}}]{Schlegel}
{Schlegel} D.~J., {Finkbeiner} D.~P., {Davis} M., 1998, \apj, 500, 525

\bibitem[{{Skrutskie} {et~al}\mbox{.}(2006){Skrutskie}, {Cutri}, {Stiening},
  {Weinberg}, {Schneider}, {Carpenter}, {Beichman}, {Capps}  \&
  {Chester}}]{2MASS}
{Skrutskie} M.~F. {et~al.}, 2006, \aj, 131, 1163

\bibitem[{{Stark} \& {Wade}(2003)}]{Stark2003}
{Stark} M.~A., {Wade} R.~A., 2003, \aj, 126, 1455

\bibitem[{{Stroeer} {et~al}\mbox{.}(2007){Stroeer}, {Heber}, {Lisker},
  {Napiwotzki}, {Dreizler}, {Christlieb}  \& {Reimers}}]{Stro2007}
{Stroeer} A., {Heber} U., {Lisker} T., {Napiwotzki} R., {Dreizler} S.,
  {Christlieb} N., {Reimers} D., 2007, \aap, 462, 269

\bibitem[{{Ulla} \& {Thejll}(1998)}]{Ulla98}
{Ulla} A., {Thejll} P., 1998, \aaps, 132, 1

\bibitem[{{Vennes} {et~al}\mbox{.}(2011){Vennes}, {Kawka}  \&
  {N{\'e}meth}}]{Vennes2011}
{Vennes} S., {Kawka} A., {N{\'e}meth} P., 2011, \mnras, 410, 2095

\bibitem[{{Vos} {et~al}\mbox{.}(2013){Vos}, {{\O}stensen}, {N{\'e}meth},
  {Green}, {Heber}  \& {Van Winckel}}]{Vos13}
{Vos} J., {{\O}stensen} R.~H., {N{\'e}meth} P., {Green} E.~M., {Heber} U., {Van
  Winckel} H., 2013, \aap, 559, A54

\bibitem[{{Wright} {et~al}\mbox{.}(2010){Wright}, {Eisenhardt}, {Mainzer},
  {Ressler}, {Cutri}, {Jarrett}, {Kirkpatrick}, {Padgett}, {McMillan},
  {Skrutskie}, {Stanford}, {Cohen}, {Walker}, {Mather}, {Leisawitz}, {Gautier},
  {McLean}, {Benford}, {Lonsdale}, {Blain}, {Mendez}, {Irace}, {Duval}, {Liu}
  \& {Royer}}]{WISE}
{Wright} E.~L. {et~al.}, 2010, \aj, 140, 1868

\bibitem[{{Zacharias} {et~al}\mbox{.}(2013){Zacharias}, {Finch}, {Girard},
  {Henden}, {Bartlett}, {Monet}  \& {Zacharias}}]{UCAC4}
{Zacharias} N., {Finch} C.~T., {Girard} T.~M., {Henden} A., {Bartlett} J.~L.,
  {Monet} D.~G., {Zacharias} M.~I., 2013, \aj, 145, 44

\bibitem[{{Zhang} \& {Jeffery}(2012)}]{Zhang2012}
{Zhang} X., {Jeffery} C.~S., 2012, \mnras, 419, 452

\end{thebibliography}

\appendix

\section{The SVO hot subdwarf archive}
\label{sec:apen}

In order to help the astronomical community on using the catalogue of subdwarfs identified in this paper we have developed an archive system that can be accessed from a Web page\footnote{\url{http://svo2.cab.inta-csic.es/vocats/hsa/}} or through a Virtual Observatory ConeSearch\footnote{Try for instance \url{http://svo2.cab.inta-csic.es/vocats/hsa/cs.php?RA=0&DEC=0&SR=100&VERB=2}}

\subsection{Web access}
\begin{figure*}
\resizebox{\hsize}{!}{\includegraphics[width= 0.8 \textwidth]{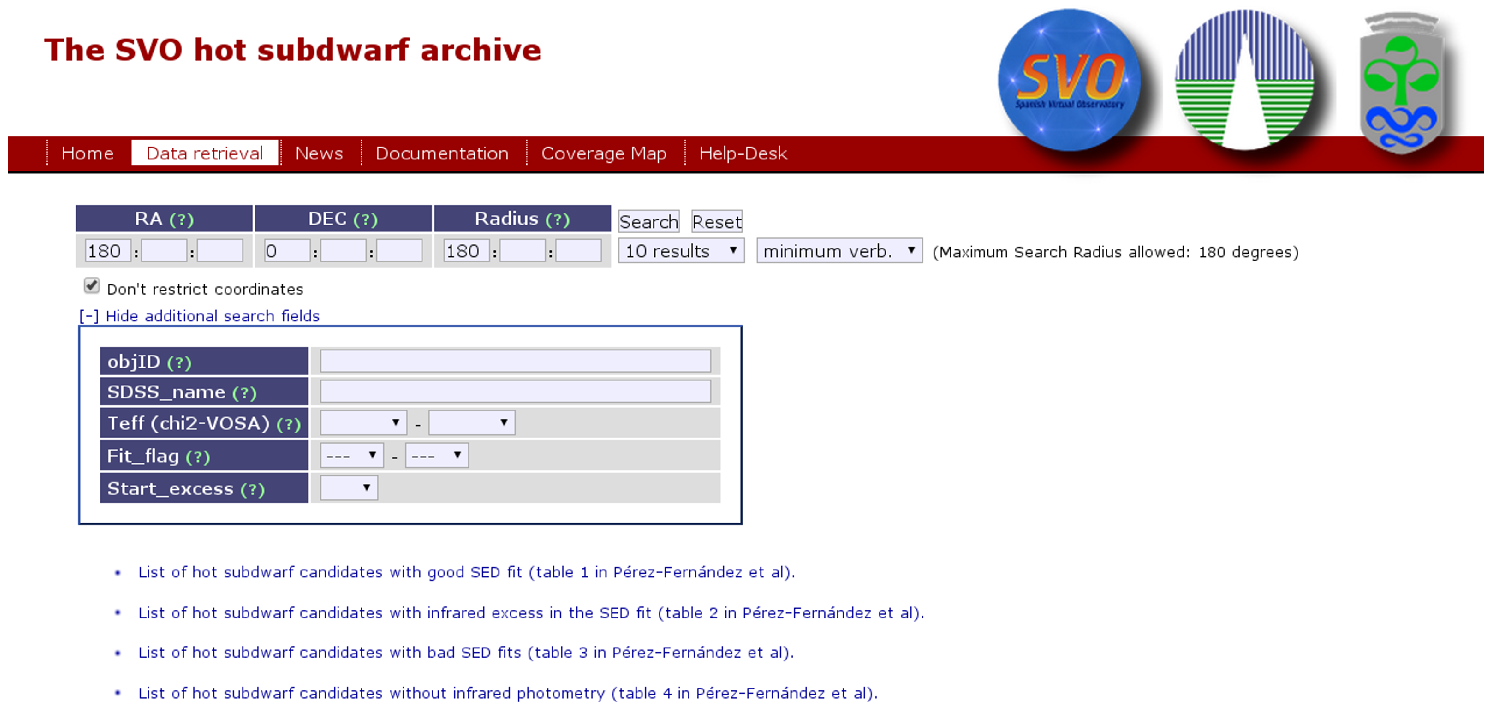}}
\caption{Web interface. Input query form}\
\label{figapen1}
\end{figure*}

The archive system implements a very simple search interface that permits queries by coordinates/radius as well as by other criteria of interest (object identifier, Teff, quality flag or excess). A selection of the astrometric, photometric and physical parameters to be displayed in the table of results can also be done assigning a type of verbosity: minimum, medium or maximum. The default search radius is set to 5 arcsec. The user can also select the maximum number of sources to return (with values ranging from 10 to unlimited) (Fig. \ref{figapen1}).

\begin{figure*}
\resizebox{\hsize}{!}{\includegraphics[width= 0.8 \textwidth]{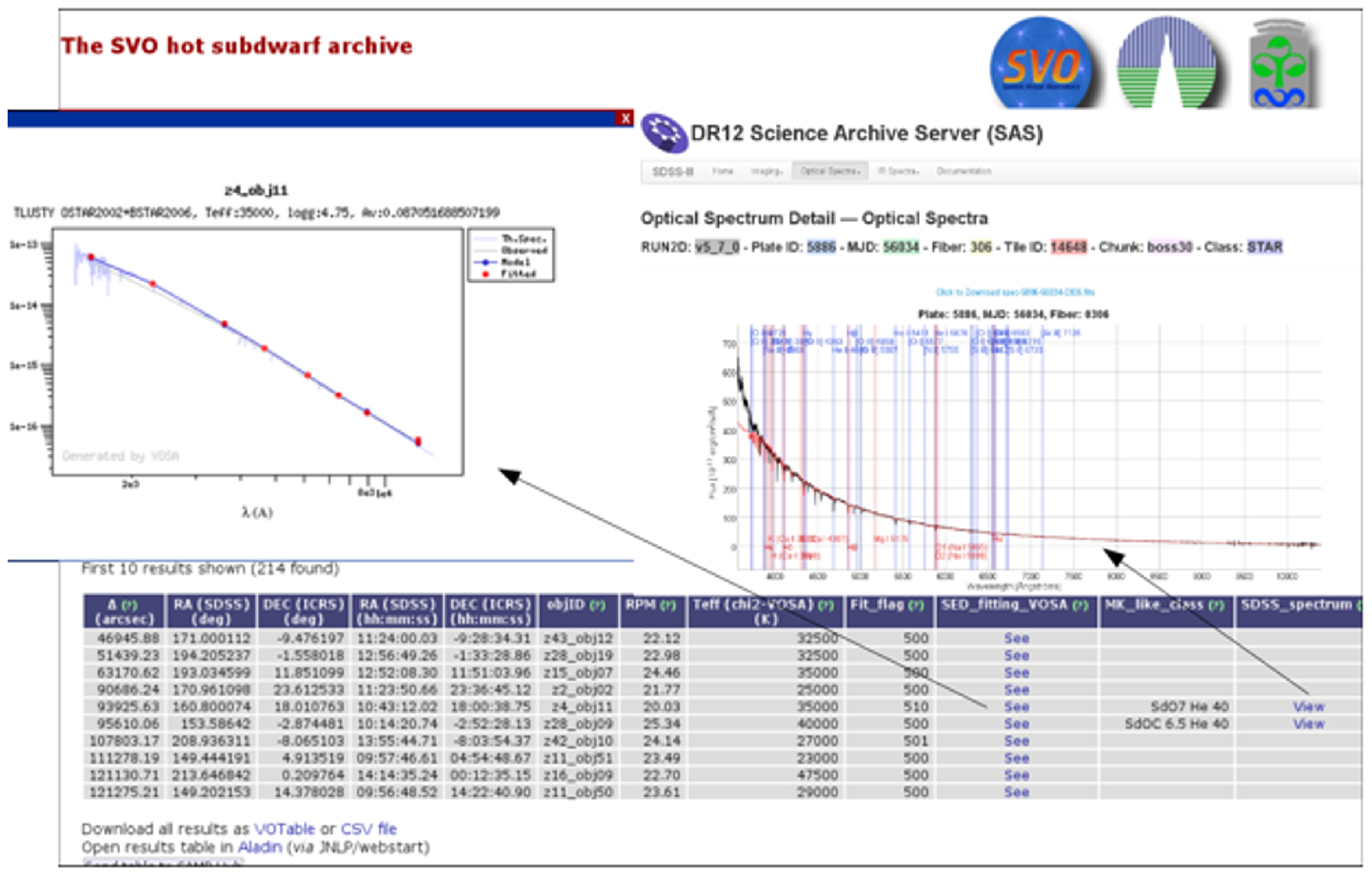}}
\caption{Result from a query. The SED and SDSS spectrum visualisation capabilities are also shown}  \label{figapen2}
\end{figure*}

The result of the query is a HTML table with all the sources found in the archive fulfilling the search criteria. Detailed information on the output fields can be obtained placing the mouse over the name of the column. The archive implements the SAMP (Simple Application Messaging Protocol). SAMP allows applications to communicate with each other in a seamless and transparent way for the user. This way, the results of a query can be easily transferred to other VO application, such as Topcat (Fig \ref{figapen2}).

\subsection{Virtual Observatory access}

The Virtual Observatory (VO)\footnote{\url{http://www.ivoa.net}} is an international initiative whose primary goal is to provide an efficient access and analysis of the information hosted in astronomical archives and services. Having a VO-compliant archive is an important added value for an astronomical project to guarantee the optimum scientific exploitation of their datasets.

Our archive system has been designed following the IVOA standards and requirements. In 
particular, it implements the Cone Search protocol, a standard defined for retrieving records from a catalogue of astronomical sources. The query made through the Cone Search service describes a sky position and an angular distance, defining a cone on the sky. The response returns a list of astronomical sources from the catalogue whose positions lie within the cone, formatted as a VOTable. 
%(Fig \ref{figapen3}).

%\begin{figure*}\resizebox{\hsize}{!}{\includegraphics[width= 0.8 \textwidth]{figure3ce.eps}}\caption{Result (in VOTable format) of a query using the Cone Search service.}  \label{figapen3}\end{figure*}

%\bsp

\label{lastpage}

\end{document}